\documentclass[a4paper,11pt]{article}
\usepackage{times}

\usepackage{graphicx}
\usepackage{algorithm}
\usepackage{algpseudocode}
\usepackage{amsfonts}
\usepackage{amsmath, amsthm, amssymb}
\usepackage{ulem}
\usepackage{color}
\usepackage{geometry}
\usepackage{hyperref}
\usepackage{ifthen}
\usepackage{pdfsync}
\usepackage{subfigure}
\usepackage[show]{ed}
\usepackage{textcomp}
\usepackage{authblk}
\title{Relative entropy minimizing noisy non-linear neural network to approximate stochastic processes}
\author[1]{Mathieu N Galtier}
\author[2,3]{Camille Marini}
\author[4]{Gilles Wainrib}
\author[1]{Herbert Jaeger}
\affil[1]{School of Engineering and Science, Jacobs University Bremen gGmbH, 28759 Bremen, Germany}
\affil[2]{Institut f\"{u}r Meereskunde, Zentrum f\"{u}r Meeres- und Klimaforschung, Universit\"{a}t Hamburg, Hamburg, Germany}
\affil[3]{MINES ParisTech, 1, rue Claude Daunesse, F-06904 Sophia Antipolis Cedex, France}
\affil[4]{Laboratoire Analyse G\'{e}om\'{e}trie et Applications, Universit\'{e} Paris XIII, France}

\def \eps {\epsilon}

\def \Q {\mathbf{Q}}

\def \E {\mathbb{E}}

\def \B {\mathbf{B}}
\def \Q {\mathbf{Q}}

\def \S {\mathbf{S}}

\def \R {\mathbb{R}}

\def \vv {\mathbf{v}}
\def \vc {\mathbf{v}^1}
\def \vr {\mathbf{v}^0}
\def \vcu {\u^{1}}

\def \u {\mathbf{u}}
\def \x {\mathbf{x}}
\def \y {\mathbf{y}}

\def \J {\mathbf{J}}
\def \W {\mathbf{W}}
\def \Wrr {\mathbf{W}_{00}}
\def \Wrc {\mathbf{W}_{01}}
\def \Wcc {\mathbf{W}_{11}}
\def \Wcr {\mathbf{W}_{10}}

\def \eqdef {\stackrel{def}{=}}

\begin{document}
\maketitle
\begin{abstract}
  A method is provided for designing and training noise-driven
  recurrent neural networks as models of stochastic processes. The
  method unifies and generalizes two known separate modeling
  approaches, Echo State Networks (ESN) and Linear Inverse Modeling
  (LIM), under the common principle of relative entropy minimization.
  The power of the new method is demonstrated on a stochastic
  approximation of the El Ni\~no phenomenon studied in climate
  research.
\end{abstract}

\section{Introduction}

Blackbox modeling methods for stochastic systems have a broad range of
applications in physics, biology, economy or the social sciences.
Generally speaking, a model of a stochastic system is a representation
of the conditional distribution of the system's future given the
present state (Markov models) or some part or the entire system past.
There is a large variety of such stochastic predictors among which we
focus on generic methods which do not depend on the type of data
considered. The Auto-Regressive-Moving-Average (ARMA) models
\cite{box2013time} form a class of linear stochastic approximators
which has led to many derivative works and is widely used in
engineering applications. In particular, it covers the case of
multivariate linear stochastic differential equations (SDE), or
Ornstein-Uhlenbeck processes, which is the basic structure used in the
Linear Inverse Modeling (LIM) theory \cite{penland1993prediction}.
ARMA models are generally learnt by optimizing a least squares measure
of the prediction error. A notable characteristic of ARMA models is
that the dimension of the underlying SDE is identical to the
observable dimension of the target time series. By contrast, dynamic
Bayesian networks \cite{murphy2002dynamic}, with Hidden Markov Models
(HMM)\cite{baum1966statistical, rabiner1989tutorial} as the most
widely employed special case, rely on hidden variables.  HMM are
trained by maximum likelihood schemes, typically with some version of
the expectation maximization algorithm
\cite{dempster1977maximum,moon1996expectation}. A problem with dynamic
Bayesian networks, inherited from their simpler static counterparts,
is that inference (e.g. prediction) quickly becomes
computationally expensive when the dependency structure of hidden
variables is not particularly simple (as it is in HMMs).  The Temporal
Restricted Boltzmann Machine \cite{HintonSutskever06}, a recent
addition to the spectrum of such models, is a point in case.  With the
advent of kernel machines in machine learning community, models based
on Gaussian Processes have been designed to approximate stochastic
processes \cite{rasmussen2006gaussian}. A critical point regarding
these models lies in their computational complexity when working with
long time series. There is also a large body of literature about
online adaptive predictors, e.g. Kalman filters
\cite{haykin2005adaptive}. In this paper however we focus on
non-adaptive models trained on all available training data using a
batch algorithm. 

Recurrent neural networks (RNNs) have also been used in various ways
for approximating stochastic dynamical systems. In their basic forms
\cite{williams1995gradient,pearlmutter1995gradient}, RNNs are models
of deterministic dynamical systems; if trained on data sampled from
stochastic sources, at exploitation time such RNNs will not propose
future distributions but only a single expected mean future
trajectory. RNNs represent, in principle, a promising model class
because they are dense in interesting classes of target
systems, implying that arbitrarily accurate models can in principle be
found \cite{funahashi1993approximation,sontag1997recurrent}.
Gradient-descent based learning algorithms for RNNs are typically
computationally expensive and cannot be guaranteed to converge. Since
about a decade, an alternative approach to RNN design and training,
now generally called \emph{reservoir computing}
\cite{jaeger2004harnessing,maass2002real}, has overcome the problem of
learning complexity. The key idea in this field is not to train all
parameters of an RNN but only the  weights of connections leading from
the RNN ``body'' (called \emph{reservoir}) to the output neurons. Here
we will build on a particular instantiation of reservoir computing
called \emph{Echo State Networks} (ESNs). 

Although deterministic models at the outset, neural network
architectures for predicting future distributions have been variously
proposed \cite{HusmeierTaylor97, buesing2011neural}, or neural
networks were embedded as components in hybrid models of stochastic
systems \cite{KroghRiis99,chatzis2011echo}. Here we propose a novel
way to use  RNNs in a stochastic framework based on the way
stochasticity is taken into account in LIM. LIM consists in tuning
both the drift and the diffusion term of an Ornstein-Uhlenbeck process
to approximate a stochastic process. First, the drift is optimized to
approximate the time series as if it were deterministic; then, the
diffusion is chosen so that the variances of both systems are
identical. LIM is widely used in climate research and stands as a
simple approach giving relatively good results
\cite{penland1996stochastic,hawkins2011evaluating, zanna2012forecast,
  barnston2012skill, newman2013empirical}. 

To compare two stochastic processes, and thus to define what it means
to approximate a stochastic process, we use the relative entropy (also
known as Kullback-Leibler divergence) \cite{kullback1951information}. Although not a true distance, it displays many interesting properties, interpretations and relationships with other quantities such as the mutual information \cite{cover2012elements} or the rate function in large deviations theory \cite{ellis2005entropy}. It also is computationally convenient (as opposed to the Wasserstein distance for instance), and
has been widely used used in machine learning \cite{ackley1985learning,hinton2006fast}.
Usually, this measure is used to compare the laws of two discrete or continuous random variables, but it can also be used to compare the laws of two stochastic processes in the path space, which is at the basis of this paper. This way of measuring the difference in law between two stochastic processes amounts in performing a change of probabilities thanks to Girsanov Theorem \cite{karatzas1991brownian}, whose applications range from mathematical finance \cite{avellaneda1997calibrating} to simulation methods for rare events \cite{wainrib2013some}. In the context of recurrent neural networks, we have already
shown that the learning rule deriving from the minimization of relative entropy has
interesting biological features since it combines two biologically plausible learning mechanisms \cite{galtier2013biological}

In this paper, we show how to train a noise-driven RNN to minimize its
relative entropy with respect to a target process. The method consists
two steps. First, the drift of the neural network is trained by
minimizing its relative entropy with respect to the target (Section
\ref{sec: drift}). Second, the noise matrix of the network is
determined based on a conservation principle similarly to LIM (Section
\ref{sec: noise}). We show how this approach extends the existing ESN
and LIM theory in Section \ref{sec: comparison}. Numerical
approximations to the double well potential and to the El Ni\~{n}o
phenomenon studied in climate research are presented in Section
\ref{sec: simus}.

\section{Model}\label{sec: model}
We define here two mathematical objects that are, a priori, unrelated:
a stochastic time series and an autonomous RNN made of two layers. The
time series is assumed to be a sample path of an underlying stochastic
process which is the modelling target. The objective is to make the
RNN approximate the target process.

The target time series $\u$ is assumed to be the discretization of an
$n$-dimensional ergodic continuous process defined on the time interval
$[0,T]$. The discretization step is chosen to be $dt = 1$ which
corresponds to fixing the timescale. Imposing $T \in \mathbb{N}$, $\u$
can be seen as a matrix in $\R^{n \times T}$. For each $t \in
\{1,..,T\}$, we use the notation $\u_t$ for the $n$-dimensional vector
corresponding to the value of the continuous target time series at
time $t$. Similarly, we write $\delta \u_t \eqdef \u_{t+1} - \u_{t}$
and $\delta{\u} \in \R^{n \times T}$ corresponding to the previous
definition (with the convention that $\delta{\u}_T = 0$).

\begin{figure}[!ht]
 \centering
 \includegraphics[width=0.5\textwidth]{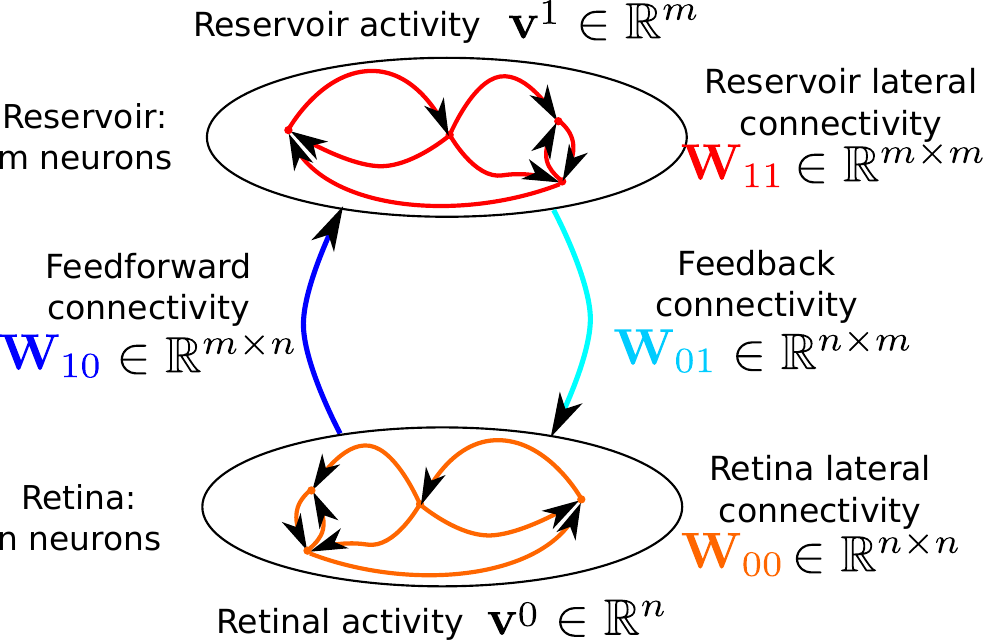}
 \caption{Structure and main notations of the neural network described in Section \ref{sec: model}.}
 \label{fig: neural network}
\end{figure}

The two-layer neural network is defined as follows. The first layer,
also called retina, has $n$ neurons, as many as the target time series
dimension. We take $\vr_t \in \R^n$ to be the activity of the retina
at time $t$ which will eventually approximate the target time series.
The second layer, also called reservoir, has $m$ neurons. Because each
reservoir neuron does not directly correspond to a variable of the
target, they are said to be hidden neurons. We denote the activity of
the reservoir at time $t$ by $\vc_t \in \R^m$. Each layer has a
complete internal connectivity, recurrent connections, that is, all
neurons within a layer are interconnected. The two layers are
interconnected with feedforward, i.e. retina to reservoir, connections
and feedback, i.e. reservoir to retina, connections, as shown in
Figure \ref{fig: neural network}.a.  In this paper, according to a
guiding principle in reservoir computing \cite{LukoseviciusJaeger09}, the feedforward and
reservoir matrices $\Wcr$ and $\Wcc$ are drawn randomly and remain
unchanged. Only connections leading to retina neurons will be adapted.
These are collected in $\W = (\Wrr\ \Wrc) \in \R^{n \times (n + m)}$.

The activity of each layer is governed by the following differential law:
\begin{equation}
\left\{
 \begin{array}{cl}
  d\vr_t = & \big(-\vr_t + \Wrr\vr_t + \Wrc\vc_t) dt  + \Sigma dB_t\\
  d\vc_t = & \eps\big(-l\vc_t + s(\Wcr\vr_t + \Wcc\vc_t)\big) dt
 \end{array}
\right.
\label{eq: activities}
\end{equation}
where $\eps, l\in \R_+$, $s$ is a sigmoid function, e.g. $\tanh$, that
is applied elementwise, i.e. $s(\x)_i = s(\x_i)$, $\Sigma  \in \R^{n
  \times n}$ is the noise matrix and $B_t$ is an $n$-dimensional
Brownian motion.

In order to  unify LIM and ESNs, we submit this architecture to
certain restrictions.  In particular, we choose the first
layer to be linear and we choose a $\tanh$ nonlinearity in the
reservoir. Later, we will make a  simple choice for a numerical
differentiation scheme for the same reason. 

\section{Training to  minimize relative entropy}\label{sec: drift}
This section explains the training of the connection matrices $\Wrr$ and
$\Wrc$ so that the distance between the neural network and the
target time series is minimized. In other words, the drift of the
neural network will be designed to match that of the target stochastic
process.

\subsection{Relative entropy between target and retina}\label{sec: entropy}
We now define a quantity measuring the dynamical distance between the
target time series and the retinal activity. At first sight, these two
mathematical objects have a different nature: the first is a time
series and the second is a dynamical system. However, we assume  that
there exists a dynamical system (possibly very complicated
and/or with hidden variables) which has generated the target
time series. Thus, we want to compute the distance between this system
and the neural network. A natural measure of similarity between
stochastic processes is the relative entropy or Kullback-Leibler
divergence. First, we show how to compute the relative entropy between
two diffusion processes with the same diffusion. Second, we apply it
to computing the ``distance`` between the neural network and the
target time series.

\subsubsection{Relative entropy between two diffusion processes sharing the same diffusion coefficient}\label{sec: entrop diffusion proc}
We now introduce the computation of the relative entropy, or Kullback-Leibler divergence, between two $n$-dimensional diffusion
processes $\x$ and $\y$ sharing the same diffusion coefficient $\Sigma  \in \R^{n \times n}$. 
Let $\{\Omega, \mathcal{F}, P, \{\mathcal{F}_t\}_{0\leq t\leq T}\}$ be a probability space, equipped with the natural filtration of the standard Brownian motion. Consider two diffusion processes $(\x_t)$ and $(\y_t)$ in $\R^n$ under $P$, solutions of the following stochastic differential equation:
\begin{equation}
d\x_t = f(\x_t)dt + \Sigma dB_t
\label{eq: stoc proc 1}
\end{equation}
\begin{equation}
d\y_t = g(\y_t)dt + \Sigma dB_t
\label{eq: stoc proc 2}
\end{equation}
where $B_t$ is a n-dimensional $P$-Brownian motion.

Defining and computing the relative entropy naturally follows from the
application of the Girsanov theorem \cite{girsanov1960transforming}
described in chapter 3.5 of Karatzas and Shreve's textbook
\cite{karatzas1991brownian}\footnote{we apply the theorem 5.1 with, in
  their notations, $X_t = \Sigma ^{-1}\big(g(\x_t) - f(\x_t)\big)$,
  $W_t = B_t$, $\tilde{W}_t = \tilde{B}_t$ and $Z_T$ is the
  Radon-Nikodym derivative of $Q$ with respect to $P$ according to
  equation 5.4.}. It provides a stochastic change of variable which
makes it possible to change the drift of a diffusion process provided
the underlying measure of the Brownian motion is changed accordingly.
Indeed, there exists a probability measure $Q$ and $\tilde{B}_t$ a
$n$-dimensional $Q$-brownian motion, such that the stochastic process
$(\x_t)$ is also solution of
\begin{equation}
d\x_t = g(\x_t)dt + \Sigma d\tilde{B}_t
\label{eq: stoc proc transformed}
\end{equation}
The probability measure has to be coherent with this change of drift,
which is enforced through the Radon-Nikodym derivative of $Q$ with
respect to $P$
\begin{equation}
\frac{dQ}{dP} = \exp\left(\sum_{i=1}^n\int_0^T \Big[\Sigma ^{-1}\big(g(\x_t) - f(\x_t)\big)\Big]_i dB_{t,i} - \frac{1}{2} \int_0^T\Big\|\Sigma ^{-1}\big(f(\x_t) - g(\x_t)\big) \Big\|^2 dt\right)
\label{eq: Radon-Nikodym}
\end{equation}
where $\Big[\Sigma ^{-1}\big(g(\x_t) - f(\x_t)\big)\Big]_i$ and
$dB_{t,i}$ are the $i^{th}$ components of $\Sigma ^{-1}\big(g(\x_t) -
f(\x_t)\big)$ and $dB_{t}$ respectively. 

For this quantity to be well-defined, the probability measure $P$ has to be absolutely continuous with respect to $Q$. This is a consequence of the technical condition $ \int_0^T\E[\|\Sigma ^{-1}\big(f(\x_t) - g(\x_t)\big)\|^2] dt< \infty$. 

Given that both processes $(\x_t)$ and $(\y_t)$ are written with the same drift \eqref{eq: stoc proc 2}, \eqref{eq: stoc proc transformed}, it is natural to consider the relative entropy between the processes as the relative entropy between the measures $P$ and $Q$ which reads 
\begin{equation}
 H(\y | \x) \eqdef H(Q | P) \eqdef \E_Q\left[\ln\Big(\frac{dQ}{dP}\Big) \right]
\end{equation}
Using \eqref{eq: Radon-Nikodym} leads to
\begin{equation}
 H(\y | \x) = \frac{1}{2} \E\left[\int_0^T\Big\|\Sigma ^{-1}\big(f(\x_t) - g(\x_t)\big)\Big\|^2 dt \right]
 \label{eq: relative entropy sto}
\end{equation}

Strictly speaking, this quantity is not a distance since it is not
symmetric. Yet it is always positive and zero only when the two drifts
$f$ and $g$ are equal. This makes it a natural and useful measure of
the similarity of two stochastic processes.

Note that although the drift $g$ initially corresponds to the
stochastic process $(\y_t)$, it is evaluated at the value $\x_t$.
This is one of the main feature of relative entropy: it
measures the difference of the drifts along the trajectory of one
process.

It is crucial that the diffusion matrix $\Sigma $ is the same in both equations \eqref{eq: stoc proc 1} and
\eqref{eq: stoc proc 2}.  If it were not the case, the two
measures $P$ and $Q$ would  not be absolutely continuous and the
Radon-Nikodym derivative, and a fortiori the relative entropy, would not
be defined.

\subsubsection{Application to our case}\label{sec: discretizing relative entropy}
The main conceptual problem to apply the previous result is that, in
practice, we do not know the continuous-time stochastic process which
we assume has generated the discrete target time series. The time series is the only
piece of information we have. Actually, many stochastic processes could have generated
this discrete time series. We want to find the stochastic processes of the form \eqref{eq: stoc
  proc 1}, which are the more likely to have produced the time series. 
  Let us call $f$ the smooth function that defines the
diffusion process \eqref{eq: stoc proc 1} which was most likely to produce
the target time series. Although we do not and will not know the
explicit formula for $f$, we formally define the distance between the neural
network \eqref{eq: activities} and the target time series as the
relative entropy between the neural network and the diffusion process
\eqref{eq: stoc proc 1} with this particular $f$. We will eventually make this quantity computable.

We also need to bridge the gap between the discrete definition of the
target time series and the continuous formulation of the relative
entropy in \eqref{eq: relative entropy sto}. This can be done
by discretizing equation \eqref{eq: relative entropy sto} on the
partition adapted to the definition of $\u$. Assuming that the
sampling of $\u$ is fine enough, we can reasonably replace the integral
by a discrete sum. At this step we can also use the ergodicity
property of the target time series to drop the expectation in this
equation. Therefore, a first tentative of definition of the relative
entropy between the target time series and the neural network is
\begin{equation}
 H = \frac{1}{2 T} \sum_{t=0}^T \Big\|\Sigma ^{-1}\big(f(\u_t) - g(\u_t)\big)\Big\|^2 + O(T^{-1/2})
 \label{eq: discretized entropy}
\end{equation}
The term $O(T^{-1/2})$ accounts for the
ergodicity approximation when the total time window $T$ is not
infinite.

However, in practice, it is not possible to have a direct access to
$f(\u)$. Estimating the drift $f(\u)$ from the observation of the
time-series $\u$ belongs to the class of problems called
\textit{numerical differentiation}. Since the seminal work of Savitzky
and Golay \cite{savitzky1964smoothing}, in which the signal is first
approximated by moving polynomials before differentiation, a large
number of numerical methods have been introduced, from finite
difference methods to regularization or algebraic methods (see
\cite{liu2011error} and references therein). However, for simplicity
and to rigorously relate to ESNs and LIM, we will simply approximate
$f(\u)$ by a \textit{temporal difference} approximation which we
called $\delta \u$. Recall
\begin{equation}
 \delta \u_t = \u_{t+1}-\u_t = \left(\int_t^{t+1} f(\x_s)ds + \Sigma  \xi_t\right)
\end{equation}
where $\x_t$ is the realization of the process from which $\u$ was
sampled and the $\xi_t$ are i.i.d standard Gaussian random variables.
We now assume that $f$ is smooth enough and that the sampling of the
stochastic process realization, which leads to defining the time
series, is fine enough, so that the following approximation (of order
0) holds:
\begin{equation}
 \int_t^{t+1} f(\x_s)ds \simeq f(\u_t)
 \label{eq: hypothesis good sampling}
\end{equation}

We now form the difference $\|\Sigma ^{-1}\big(f(\u_t)-g(\u_t)\big)\|^2$ and introduce $\Sigma ^{-1}\delta \u_t$ in the computation. This leads to
\begin{eqnarray*}
 \|\Sigma ^{-1}\big(f(\u_t)-g(\u_t)\big)\|^2&\simeq& \|\Sigma ^{-1}\big(f(\u_t)-\delta \u_t\big)\|^2 \\
 &+& 2\langle \Sigma ^{-1}\big(f(\u_t)-\delta \u_t\big), \Sigma ^{-1}\big(\delta \u_t - g(\u_t)\big)\rangle  \\
 &+& \|\Sigma ^{-1}\big(\delta \u_t - g(\u_t)\big)\|^2\\
 &\simeq& \|\xi_t\|^2 \\
 &+&  2\Big \langle \xi_t, \Sigma ^{-1}f(\u_t) + \xi_t \Big\rangle \\
 &-& 2\langle  \xi_t,\Sigma ^{-1}g(\u_t) \rangle\\
 &+&  \|\Sigma ^{-1}\big(\delta \u_t - g(\u_t)\big)\|^2
  \end{eqnarray*}
Let us look at what becomes each of the four lines above when taking the empirical average (or equivalently the expectation by ergodicity):
\begin{itemize}
\item The first line becomes $1$.
\item In the second line, the first term $2\langle \xi_t, \Sigma ^{-1}f(\u_t)\rangle$ has zero mean and thus vanishes. The second term becomes $2$.
\item The third line is centered and thus vanishes.
\item Finally, the last term is the one we want to keep in the algorithm. 
\end{itemize}

To summarize, it remains
\begin{equation}
H \simeq \frac{1}{2 T}\Big(\sum_{t=0}^T  \big\|\Sigma ^{-1}\big(\delta \u_t-g(\u_t)\big)\big\|^2\Big) + O(T^{-1/2})
\label{eq: relative entropy with errors}
\end{equation}

One could also use a Taylor expansion for equation \eqref{eq:
  hypothesis good sampling} which would add some corrective terms to
\eqref{eq: relative entropy with errors}. However, provided that the
sampling of the target is fine enough, these terms can be neglected.

To be fully exhaustive, one should also look at the term $O(T^{-1/2})$
coming from the central limit theorem for ergodic convergence, and
this term may also have some contribution which depends on the
connectivity, in particular from the second term in line 3: $2\langle
\xi_t,\Sigma ^{-1}g(\u_t) \rangle$. It is zero-mean, so in the limit
$T\to \infty$ it will disappear when we take the empirical average,
but its variance is in fact $4 \E[\Sigma ^{-1}g(\u)^2]$ which is not
necessarily zero.

One last modification of the formal definition in equation \eqref{eq:
  relative entropy sto} stems from its problematic dependence on the
noise matrix $\Sigma $. Recall we intend to tune the noise to
minimize the relative entropy. Given the definition \eqref{eq:
  relative entropy sto} a trivial choice is to have an extremely
strong noise: in this case the two stochastic processes are so random that they can be said to be identical. However,
this is a situation which we would like to avoid: we want to find a
reasonable noise matrix $\Sigma $. To correct this pathologic behavior,
it seems natural to divide the formal definition \eqref{eq: relative
  entropy sto} by the square of the operator norm of $\Sigma ^{-1}$
which we write $|||\Sigma ^{-1}|||^2$ (or equivalently multiplying by
the square of the smallest eigenvalue of $\Sigma $). The resulting
quantity simply is proportional to the original definition, but
without this problem.  Note that this does not mean that we restrict
our approach to unit norm noises matrices as is shown later.

This finally leads to a definition corresponding to equation \eqref{eq: relative entropy with errors} without the various error terms. Besides, we also want to take into account a regularization term.  This leads to defining an analog to regularized relative entropy between the target time series $\u$ and our neural network \eqref{eq: activities} (respectively corresponding to $\x$ and $\y$ in \eqref{eq: stoc proc 1} and \eqref{eq: stoc proc 2}) as
\begin{equation}
 H_\u(\W) \eqdef \frac{1}{2 T}\sum_{t=0}^T \Big\|\S^{-1}\big(-\u_t + \Wrr\u_t + \Wrc\vcu_t - \delta \u_t\big)\Big\|^2 + \frac{\alpha^2}{2} \|\S^{-1}\W\|^2
 \label{eq: relative entropy}
\end{equation}
where $\S^{-1} = \frac{\Sigma ^{-1}}{|||\Sigma ^{-1}|||}$ and $\vcu \in \R^{m \times T}$ is the solution of 
\begin{equation*}
\delta{\vcu_t} =  \eps\big(-l\vcu_t + s(\Wcr\u_t + \Wcc\vcu_t)\big)
\end{equation*}
governing the reservoir when it is fed by the target time series (as opposed to the retinal activity). The definition of $\delta{\vcu} \in \R^{m \times T}$ is similar to that of $\delta{\u}$. In the following, we abusively call $H_\u(\W)$ the relative entropy between the target time series and the network.

The second term in \eqref{eq: relative entropy} is a regularization
term which ensures decent generalization properties of the following
algorithm. It does not change the qualitative meaning of the notion of
relative entropy but penalizes the networks with high connection
strength. The reason why it contains a multiplication by $\Sigma ^{-1}$
will become clear later. In short, it will make the system
decoupled and we will be able to compute subsequently the connections
first and the noise matrix second.

\begin{figure}[!ht]
 \centering
 \includegraphics[width=0.4\textwidth]{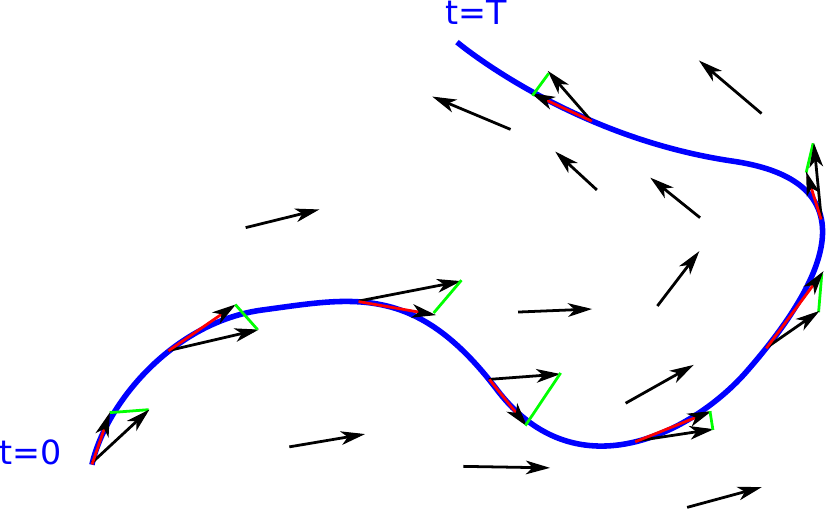}
 \caption{Illustration of the computation of the relative entropy in the case $\Sigma  = I_d$. It corresponds to integrating all the green bars along the trajectory of the target time series. The solid blue line corresponds to the target time series. The red arrows correspond to the speed vectors of the target time series. The black arrows correspond to the vector field of the retinal neural network.}
 \label{fig: relative entropy interpretation}
\end{figure}

One can also understand the definition \eqref{eq: relative entropy}
without referring to the analogy with relative entropy, in the special
case $\Sigma  = I_d$ and $\alpha = 0$. Indeed, as shown in Figure
\ref{fig: relative entropy interpretation}, it corresponds to
integrating the distance between the derivative of the target time
series and the flow of the retinal activity along the trajectories of
the target time series. Therefore, it is obvious that if this quantity
is null then the vector field of the retinal activity will be tangent
to the target time series derivative. Therefore, initializing the
network with the value of the target time series at time $t=0$, will
lead the network to reproduce precisely the target time series.

Observe that the relative entropy \eqref{eq: relative entropy} (even
more the rigorous definition \eqref{eq: relative entropy sto}) are
quite similar to the quantity usually minimized for prediction
\cite{williams1995gradient, pearlmutter1995gradient,
  bishop2006pattern}. The main difference is that this is the integral
of the distance between the \textit{derivative} of the target time
series and activity variable, instead of the mere distance between
target and activity. 

Note that the relative entropy \eqref{eq: relative entropy} can easily
be shown to be proportional to the negative of the log-likelihood of
the target given the neural network. This close relationship shows
that, in this case where the stochastic processes are diffusion
processes, minimizing the relative entropy is rigorously equivalent to
maximizing the log-likelihood.

\subsection{Gradient of the relative entropy}\label{sec: gradient}
Now, the idea is to compute the gradient of $H_\u(\W)$ with respect to
retinal and feedback connectivities. This will be useful because the gradient cancels out
at the minimum of the relative entropy. This gives us a useful way to
compute the connectivity $\W^* = (\Wrr^*\ \Wrc^*) \in \R^{n \times (n
  + m)}$ such that the network best approximates  the
target stochastic process. Because the relative entropy~\eqref{eq:
  relative entropy} is quadratic in $\W$, it is  convex. Thus, it
has a single critical point which is a global minimum.

For readability, we introduce the continuous $n$n-dimensional function
defined on $[0,T]$ by 
\begin{equation}
 \zeta(\W) =  \Wrr\u + \Wrc\vcu - \delta \u - \u
 \label{eq: zeta}
\end{equation}
such that
\begin{equation}
 H_\u(\W) = \frac{1}{2}\sum_{t=0}^T \Big\|\S^{-1}\zeta_t(\W)\Big\|^2 + \frac{\alpha^2}{2} \|\S^{-1}\W\|^2
\end{equation}

The gradient will be computed in an operator framework, i.e. as the
differential of the relative entropy. Observe that the differential is
a linear operator such that
\begin{multline}
  d_\W H_\u (\J) =  \frac{1}{2}d_\W \Big(\sum_{t=0}^T \langle \S^{-1}\zeta_t, \S^{-1}\zeta_t \rangle + \alpha^2 \langle \S^{-1}\W, \S^{-1}\W \rangle \Big)(\J) \\
  = \sum_{t=0}^T \frac{1}{2}d_\W\big\langle \S^{-1}\zeta_t, \S^{-1}\zeta_t \big\rangle (\J) + \alpha^2 \langle\S^{-1}\W, \S^{-1}\J \rangle\\
  =  \sum_{t=0}^T \langle \S^{-1}\zeta_t(\W), \S^{-1}d_\W\zeta_t(\J) \rangle+  \alpha^2 \langle\S^{-1}\W,\S^{-1} \J \rangle\\
  = \sum_{t=0}^T \langle \big(\S \S')^{-1}\zeta_t(\W),
  d_\W\zeta_t(\J) \rangle + \alpha^2 \langle(\S \S')^{-1}\W, \J
  \rangle
\label{eq: differential}
\end{multline}
Because $\zeta(\W)$ is affine in $\Wrr$ and $\Wrc$, it appears that
\begin{equation}
 d_{\Wrr}\zeta(\J) = \J\u \quad \mbox{and} \quad d_{\Wrc}\zeta(\J) = \J\vcu
\end{equation}
In both cases, we have a differential of the form  $d_\W\zeta(\J) = \J\mathbf b$, with $\mathbf b = \u$ (resp. $\mathbf b = \vcu$) is a matrix of $\R^{n \times T}$ (resp. $\R^{m \times T}$).\\
Observe that ${\nabla_\W H_\u}_{ij} = d_\W H_\u (\mathbf{E}^{ij})$
where $\mathbf{E}^{ij}$ is the canonical matrix made of zeros except
at position $ij$ where it is one. Then,
\begin{multline*}
d_\W H_\u (\mathbf{E}^{ij}) =  \sum_{t=0}^T \langle (\S \S')^{-1}\zeta_t(\W), \mathbf{E}^{ij}\mathbf b_t \rangle + \alpha^2 \langle(\S \S')^{-1}\W, \mathbf{E}^{ij} \rangle \\
= \sum_{t=0}^T \big\{(\S \S')^{-1}\zeta_t(\W)\big\}_i\mathbf b_{jt} + \alpha^2 \{(\S \S')^{-1}\W\}_{ij} \\
= \big\{(\S \S')^{-1}\zeta(\W) \mathbf b'\big\}_{ij}+ \alpha^2 \{(\S \S')^{-1}\W\}_{ij}
\end{multline*}
which leads to
\begin{equation*}
\nabla_\W H_\u =(\S \S')^{-1}\zeta(\W) \mathbf b' + \alpha^2 (\S \S')^{-1}\W
\end{equation*}
Using the definition of $\zeta$ leads to
\begin{equation}
 (\S \S') \nabla_{\W}H_\u = - 
 \Big[(\delta \u + \u) \u'\quad (\delta \u + \u) {\vcu}' \Big] + 
 (\Wrr\ \Wrc)  
 \left(\alpha^2 I_d + \left[
 \begin{array}{cc}
  \u \u' & \u {\vcu}'\\ 
  \vcu \u' & \vcu {\vcu}'
 \end{array}
 \right]
 \right)
  \label{eq: feedback and retina gradient}
\end{equation}

The global minimum can be computed as $\nabla_{\W^*}H_\u =0$. Since we have assumed that $\S$ is a full rank matrix, this leads to the following formula
\begin{equation}
\begin{array}{ccl}
 \W^* & = & 
 \Big[(\delta \u + \u) \u'\quad (\delta \u + \u) {\vcu}' \Big]
   
  \left(\alpha^2 I_d + \left[
 \begin{array}{cc}
  \u \u' & \u {\vcu}'\\ 
  \vcu \u' & \vcu {\vcu}'
 \end{array}
 \right]\right) ^{-1}  \\
 & = & (\delta \u + \u)   
 \begin{pmatrix}
  \u \\ 
  \vcu
 \end{pmatrix}'
  
 \left(\alpha^2 I_d +
  \begin{pmatrix}
   \u \\ 
   \vcu
  \end{pmatrix}
   
  \begin{pmatrix} 
   \u \\ 
   \vcu
  \end{pmatrix}'
 \right)^{-1}
 \end{array}
\label{eq: equilibrium connectivity}
\end{equation}
It is interesting to observe that the solution does not depend on the
noise matrix $\Sigma $. It is this property which makes it
possible for the problem to be decoupled in two parts: (i) computation
of $\W^*$ and (ii) computation of $\Sigma $.

\section{Computing the noise with a conservation principle}\label{sec: noise}
This section is devoted to computing the noise matrix
$\Sigma $ such that the neural network matches the statistics of the target
time series. Obviously the choice of a simple additive noise (i.e.
$\Sigma $ does not depend on $\vv$ nor $t$) in system \eqref{eq:
  activities}, restricts the class of target system the neural network
can approximate accurately. Although the match will not be perfect, we
will see the method provides a reasonable and (more crucially)
coherent noisy neural network approximating the statistics of the
target.

At first sight, it may seem appropriate to choose $\Sigma $ so that the
covariance of the retinal activity matches that of the target.
However, the non-linearity in the reservoir makes it seemingly
impossible to compute analytically the covariance of system \eqref{eq:
  activities}. Therefore, we have not been able to use this idea to
fix $\Sigma $.

Another method consists in using a generalized fluctuation dissipation
relation. Penland and Matrosova \cite{penland1994balance} have
detailed a method to use a conservation principle to link the
correlation of the activity, the flow of the retina and the matrix
$\Sigma \Sigma '$. Following the lines of their derivation, we
generalize their approach to the case of reservoirs.

The generalized fluctuation dissipation relation is based on the
Fokker-Planck equation \cite{risken1996fokker} of the neural network
\eqref{eq: activities}. Any stochastic differential system can be
described equivalently by a sample path dynamics governed by
\eqref{eq: activities} or a Fokker-Planck equation which governs the
evolution of the probability density function. It corresponds to the
Eulerian description of the original stochastic differential equation.
It can intuitively be understood as a balance of how much goes in and
out of a small box centered on $\vv$, taking into account both a drift
and a diffusion mechanism. In our case, it takes the form of the
following partial differential equation.
\begin{multline}
 \frac{\partial p(\vr,\vc,t)}{\partial t} = - \mbox{div} \left[\begin{pmatrix}-\vr + \Wrr\vr + \Wrc\vc \\ \eps\big(-l\vc + s(\Wcr\vr + \Wcc\vc)\Big)\end{pmatrix} p(\vr,\vc,t)\right]\\
 + \frac{1}{2}\Delta\left[\begin{pmatrix}\Sigma \Sigma ' & 0 \\ 0 & 0\end{pmatrix} p(\vr,\vc,t)\right]
\label{eq: Fokker-Planck}
 \end{multline}
where $\mbox{div}$ is the divergence operator, i.e. $\mbox{div}(\x) = \sum_i \frac{\partial \x_i}{\partial \vv_i}$ which corresponds to the drift, and $\Delta$ is the Laplacian operator, i.e. $\Delta \J = \sum_{i,j} \frac{\partial^2 \J_{ij}}{\partial \vv_i \partial \vv_j}$, which corresponds to the diffusion. Note that this Fokker-Planck equation is independent of the underlying choice between It\^o or Stratonovich noise in the initial system \eqref{eq: activities}, because we assumed $\Sigma $ does not depend on the activity of the network.

A conservation principle about the moments of the stochastic process $\vv$ can easily be derived from equation \eqref{eq: Fokker-Planck}. Indeed, multiply it by $\{\vr_t\}_p \{\vr_t\}_q$ (the components $p$ and $q$ of the activity $\vr_t$) and integrate over the entire domain. We can use the integration by part formula several times to get
\begin{multline}
 \frac{\partial \E[\{\vr_t\}_p \{\vr_t\}_q]}{\partial t} = \E\left[\big(-\{\vr_t\}_p + \{\Wrr\vr_t\}_p + \{\Wrc\vc_t\}_p \big)\{\vr_t\}_q\right]\\
 + \E\left[\big(-\{\vr_t\}_q + \{\Wrr\vr_t\}_q + \{\Wrc\vc_t\}_q \big)\{\vr_t\}_p\right] 
 + \Sigma \Sigma '
\label{eq: FDR components}
 \end{multline}
 
 Given that the target is ergodic, it is natural to assume that its approximation also has this property. Therefore, it is legitimate to replace the expectations by time integrals over $[0,T]$ divided by $T$ in the previous equation. In a matrix formalism, this reads
\begin{equation*}
 \frac{\vr_T {\vr_T}'-\vr_0 {\vr_0}'}{T} = \big(-\vr + \Wrr\vr + \Wrc\vc\big) {\vr}'\\
 + \vr \big(-\vr + \Wrr\vr + \Wrc\vc \big)'
 + \Sigma \Sigma '
 \end{equation*} 
 A careful inspection of the terms
 shows that the left hand side is negligible when $T$ is large enough
 (which will always be the case in practice). Thus, we will drop this
 term for simplicity (although it would not pose any problem to take
 it into account).
 
 Notice the correlations terms $\vr {\vr}'$ and $\vr {\vc}'$ in the
 previous equation. Recall our initial wish to choose $\Sigma $ so that
 both neural network and target second order moments are matched.
 Although we could not directly implement this wish, we are now able
 to replace the network correlation terms by the observed moments of
 the target in the present formulation. This ansatz leads to the
 following generalized fluctuation dissipation relation
 \cite{penland1994balance}:
\begin{equation}
 \Sigma \Sigma ' = \big(I_d - \Wrr\big) \frac{\u {\u}'}{T} - \Wrc\frac{\vcu {\u}'}{T}
 + \frac{\u {\u}'}{T}\big(I_d - \Wrr\big) - \frac{\u {\vcu}'}{T}\Wrc
\label{eq: FDR}
 \end{equation}
 This equation can be seen as a coherency requirement between drift
 and diffusion of the network and second order moments of the target.
 Fortunately, the derivation of the drift leads to an explicit
 equation \eqref{eq: equilibrium connectivity} independent of the
 matrix $\Sigma $. Therefore, the previous equation can be used to
 characterize $\Sigma $ based on the knowledge of $\W$. Given that the
 square root of matrix is not injective, there are several choices for
 the matrix $\Sigma $. They all correspond to an ambiguity on the sign
 of its eigenvalues. We arbitrarily pick one of them and have thus
 found a coherent noise matrix.
 
 \section{Comparison with existing methods}\label{sec: comparison}
Our approach takes selected features of two existing methods for
 approximation, ESN and LIM, and unifies them in a mathematical
 framework. This section is devoted to clarifying the links with these
 two methods.
 \begin{figure}[ht]
 \centering
 \includegraphics[width=0.8\textwidth]{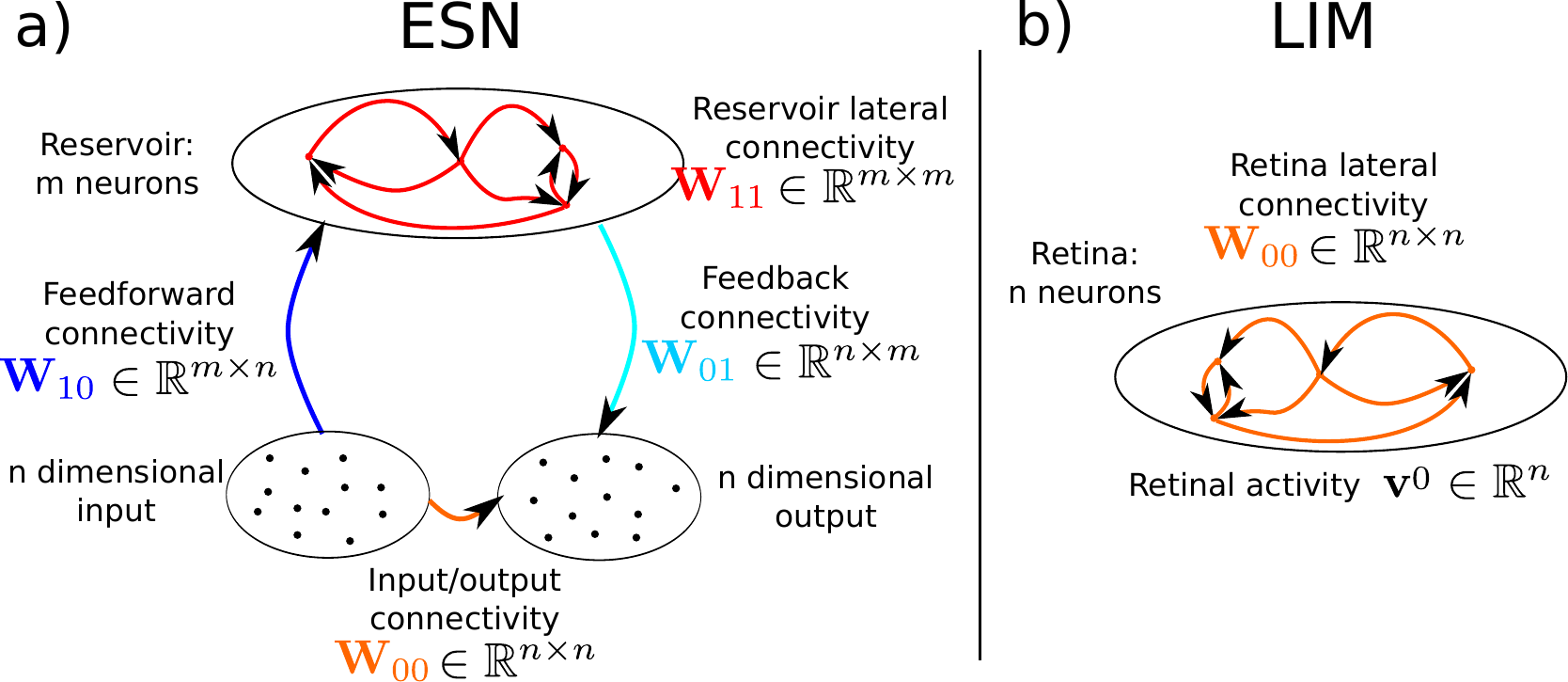}
 \caption{Architecture of related existing models. To be compared with Fig.\ref{fig: neural network}. a) Architecture of an ESN (see section \ref{sec: ESN}): the inputs and outputs are different layers. However, when the network is set to the prediction mode the value of the inputs is copied from the outputs, which corresponds to closing up the loop as in Fig.\ref{fig: neural network}. b) Architecture of LIM (see section \ref{sec: LIM}): there is no reservoir and only a retina.}
 \label{fig: architecture comparison}
\end{figure}

\subsection{Adding nonlinearities to Linear Inverse Modeling}\label{sec: LIM}
The proposed method rigorously extends LIM by adding non-linearities to the model dynamics. Linear inverse modeling consists in designing a multidimensional linear stochastic (also called Ornstein-Uhlenbeck) process which reproduces a target multivariate time series \cite{penland1995optimal, penland1996stochastic}. Naturally, the dimension of the approximating process is identical to the dimension of the target. A pervasive idea in machine learning is to consider additional (often called hidden) variables which will help the reconstruction of the target multivariate time series. In the framework presented in this paper, they correspond to the neurons in the reservoir, while the linear dynamical system analogous to LIM corresponds to the neurons in the retina. Due to the non-linearity of the dynamics of these additional variables or neurons, the present framework is as a non-linear extension of LIM. Actually, it turns out to be a surprisingly simple extension since the same formula can be used to calculate 
the linear matrix in LIM and the rectangular matrix which combines the additional variables to improve the retina's predictions.

More precisely, LIM consists in finding the matrices $\B,\Q \in \R^{n \times n}$ such that the following dynamical system reproduces the target time series $\u$.
\begin{equation*}
 \dot \vr_t = \B \vr_t + \sqrt{\Q}\dot{B}_t
\end{equation*}
where $\dot{B}_t$ is a white noise and $\sqrt{\ }$ is the matrix square root (i.e. $\Q$ is the covariance of $\sqrt{\Q}\dot{B}_t$). Note that we have intentionally used the same variable $\vr$ for the LIM and the activity in the retina.

Based on the explicit expression of the Green function of a linear system \cite{risken1996fokker}, $\B$ can be caracterized as follow
\begin{equation}
 \B = \frac{1}{\tau} \ln(\u_{.+\tau}\u'  (\u \u')^{-1})
 \label{eq: official LIM}
\end{equation}
where $\ln$ is the logarithm for matrices and $\u_{.+\tau}$ is the matrixes whose column number $t$ is $\u_{t+\tau}$ and $\tau$ is an integer usually equal to $1$ (depending on the intrinsic timescales of the target). Assuming $\tau = 1$ in the following, $\B$ can be written: 
\begin{equation*}
 \B =  \ln(I_d + (\u_{.+1} - \u) \u'  (\u \u')^{-1}) = \sum_{k=1}^{+\infty} \frac{(-1)^{k+1}}{k} \big[(\u_{.+1} - \u) \u'  (\u \u')^{-1}\big]^k
\end{equation*}
Assuming appropriate sampling of $\u$ such that $\|\delta\u\| \ll 1$, we can reasonably truncate $\B$ at first order. This leads to the following approximation: $\B \simeq \delta \u \u' (\u \u')^{-1}$.

To see the link with ESNsto, we must consider the case $m=0$ in eq.\ref{eq: activities}, so that $\W= I_d + \B$. From the above approximation, it follows $\W \simeq I_d + \delta \u \u'  (\u \u')^{-1} \simeq (\delta \u + \u)   \u'   (\u \u')^{-1}$, which corresponds to $\W^*$ in \eqref{eq: equilibrium connectivity} with $m = 0$ and $\alpha = 0$.
Given the definition of both systems, it is clear that the two methods
are identical in this restricted case. However, when the sampling is
not fine enough, equation \eqref{eq: official LIM} may lead to
solutions than differ from the log-free equation \eqref{eq: equilibrium
  connectivity}. One then could also imagine a variation of ESNs using
a matrix log when dealing with badly sampled data, but this is beyond
the scope of this paper.
 
Concerning the treatment of noise, we can observe that it is strictly 
the same with or without additional reservoir neurons.

\subsection{Adding noise to Echo States Networks}\label{sec: ESN}
The proposed algorithm rigorously generalizes ESNs to a stochastic framework. Indeed, the formula for the connectivity in equation \eqref{eq: equilibrium connectivity} is identical to the solution given by applying the classical deterministic ESN method \cite{jaeger2004harnessing,LukoseviciusJaeger09}.

To explain further the equivalence between this formalism and
classical ESNs, we now introduce the ESN formalism in its original
form as summarized in Figure \ref{fig: architecture comparison}.a. Let
us setup an ESN of reservoir size $m$ as a one time step predictor of
the input $\u$, where the input and the output have, naturally, the
same dimension $n$, and the model includes direct connections from the
input to the output. In fact, the initial setup of echo state network
makes a distinction between input and output, see Figure \ref{fig:
  architecture comparison}.a; whereas the model introduced in section
\ref{sec: model} only has a retina, see Fig.\ref{fig: neural network}.
However, ESNs can also be run in a generative mode, where the current
output becomes the next input, closing the loop between the two. This
closed loop system is precisely the same as the two layer network of
Section \ref{sec: model}, where the joined input/output nodes become the
retina with activations $\vr$, and the reservoir remains with
activations $\vc$. As summarized in Figure \ref{fig: architecture
  comparison}.a, connections from the input to the reservoir
correspond to $\Wcr$, internal reservoir connections to $\Wcc$, output
connections from the reservoir to $\Wrc$, and connections from input
directly to the output after closing the loop become the recurrent
connections in the retina $\Wrr$.

Classical ESNs are discrete-time systems, as opposed to our
continuous-time approach. Yet the two methods are closely linked: ESNs
correspond to a time-discretized version of \eqref{eq: activities}.
\begin{equation}
\left\{
 \begin{array}{cl}
  \vr_{t+1} = & \Wrr\vr_t + \Wrc\vc_t\\
  \vc_{t+1} = & s(\Wcr\vr_t + \Wcc\vc_t)
 \end{array}
\right. ,
\label{eq: esn activities}
\end{equation}
where \eqref{eq: activities} is discretized using Euler's approximation and the discretization step is taken to be equal to $1$ and $\eps=l=1$.

Training such a setup to minimize a squared error on the input prediction $\sum_{t=0}^T \left( \u_{t+1} - \vr_t \right)^2$ precisely corresponds to learning connections $\Wrr$ and $\Wrc$ according to the equation \eqref{eq: equilibrium connectivity}. Indeed, observe that $\{\delta \u + \u\}_t = \u_{t+1}$ is the prediction of the input (which corresponds to the target signal in this setup). Taking $\u$ as input and $\vcu$ as teacher-forced reservoir activations, the equation \eqref{eq: equilibrium connectivity} turns out to be the ridge regression equation generally used in \cite{LukoseviciusJaeger09}.

The treatment of noise proposed in this paper is a new contribution to the ESN theory. In that sense, this paper consists in an extension of ESNs as time series approximators to stochastic ESNs as stochastic process approximators.

\section{Numerical simulations}\label{sec: simus}
This section shows two examples of application of the proposed ESNsto
algorithm described in algorithm \ref{alg: main}.

\begin{algorithm}
\caption{ESNsto learning}
 \begin{algorithmic}
\State $\#$ Initialization:
\State Components of $\Wrc$ and $\Wcc$ are drawn randomly following a normal law. The two matrices are then rescaled get desired spectral radii.
\State $\vcu_0, \vr_0, \vc_0 \gets 0$ 
\State $\#$ Collect reservoir states:
\While{$t<T$:}
\State ${\vcu_{t+1}} =  (1 - \eps l)\vcu_t + \eps l\ s(\Wcr\u_t + \Wcc\vcu_t)$
\EndWhile
\State $\#$ Learn connections:
\State $
 (\Wrr, \Wrc)  \gets (\delta \u + \u)   
 \begin{pmatrix}
  \u \\ 
  \vcu
 \end{pmatrix}'
 \left(\alpha^2 I_d +
  \begin{pmatrix}
   \u \\ 
   \vcu
  \end{pmatrix}
  \begin{pmatrix} 
   \u \\ 
   \vcu
  \end{pmatrix}'
 \right)^{-1}
$
\State $\#$ Compute noise connections:
\State $\Sigma= \sqrt{\big(I_d - \Wrr\big) \frac{\u {\u}'}{T} - \Wrc\frac{\vcu {\u}'}{T}
 + \frac{\u {\u}'}{T}\big(I_d - \Wrr\big) - \frac{\u {\vcu}'}{T}\Wrc}$
\State $\#$ Simulate ESNsto:
\While{True}
\State $ \begin{array}{cl}
  \vr_{t+1} = & \Wrr\vr_t + \Wrc\vc_t  + \Sigma\  \mbox{randn}(n)\\
  \vc_{t+1} = & (1 - \eps l \vc_t) + \eps s(\Wcr\vr_t + \Wcc\vc_t)
 \end{array}
$
\EndWhile
\end{algorithmic}
\label{alg: main}
\end{algorithm}
It is important to realize that the
goal here is not to approximate or predict a time series, but rather
to approximate a stochastic process. Because a single stochastic
process can have different realizations, an approximation of such a
process should not aim at reproducing a given path. In this sense, we
are not dealing with classical prediction tasks and we should
exlusively focus on building a system that reproduces the law of the
target stochastic process. As a consequence, we can not compare ESNsto
with classical ESN since they do not approximate the same mathematical
objects.

We are going to compare the performances of LIM with that of
ESNsto. LIM belongs to several of the  different classes of approximators
that we mentioned in the introduction: it is a Gaussian process,
a multivariate autoregressive process of order one and an
Ornstein-Uhlenbeck process. Besides it can be easily compared to
ESNsto, since the latter generalizes the former and LIM simply is an ESNsto with 0 neurons in the reservoir. Establishing a
complete benchmark of the different methods for stochastic processes
approximation is beyond the scope of this paper. However, we point out
 the low computational complexity of learning, which is independent
of the length of the time series and mainly governed by the inversion
of a positive semi definite square matrix of size $n$. This is lower
than the complexity of Hidden Markov Models or Gaussian Processes for
long time series.

To compute the relative entropy we have used a classical
cross-validation framework. This means that we have divided the target
time series in $k$ blocks. Then for each block, we have computed the
connectivity and noise matrices on the remaining $k-1$ blocks, and
evaluated the value of the relative entropy on the selected block.
This provides a robust way to prevent  over-fitting.

This numerical section is only a proof of concept. The
(hyper)parameters of the networks (such as the spectral radius of
$\Wcc$) have been coarsely tuned, although they could significantly
improve the approximations if they were set carefully. The main reason
for our negligence is that we want
to show that an off-the-shelf ESNsto model is better than LIM and
does not require deep knowledge or large effort of neural networks
tuning.

The first example we consider is a widely considered toy model: the target is
generated by a noisy particle living in a double well. The second
example, devoted to climate modeling, will show how to approximate the
El Ni\~{n}o phenomenon in the tropical Pacific ocean.

\subsection{The noisy double-well}
The double well example explores a basic form of non-linearity. It illustrates the significant improvement brought by reservoir neurons in dealing with non-linearities.

\begin{figure}[htbp]
        \centering
                \subfigure[Double well energy landscape.]{\includegraphics[width=0.55\textwidth]{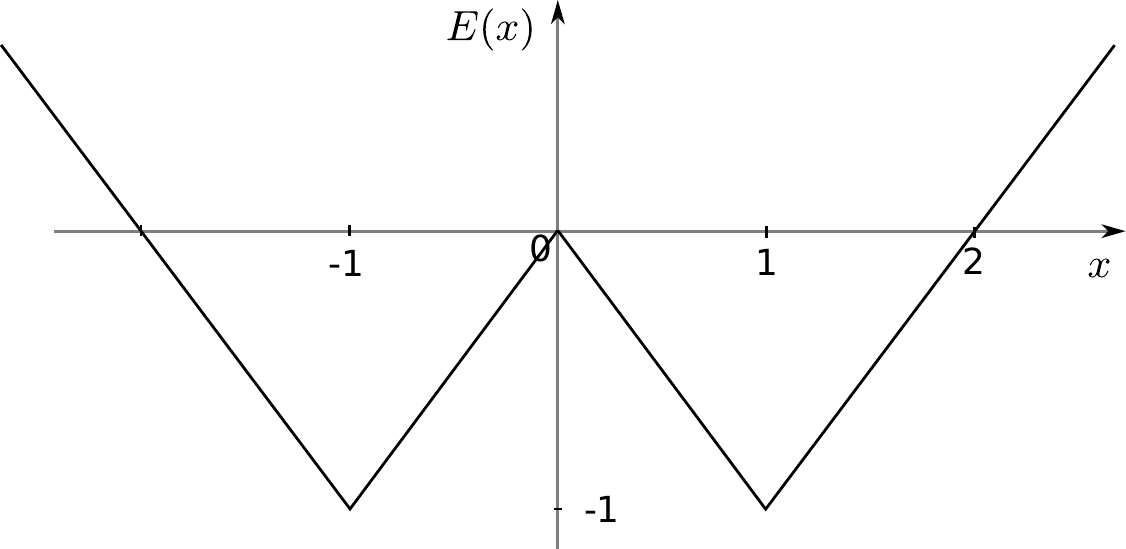}\label{fig: double well landscape}}\quad
                \subfigure[Trajectory of process \eqref{eq: double well process}.]{\includegraphics[width=0.4\textwidth]{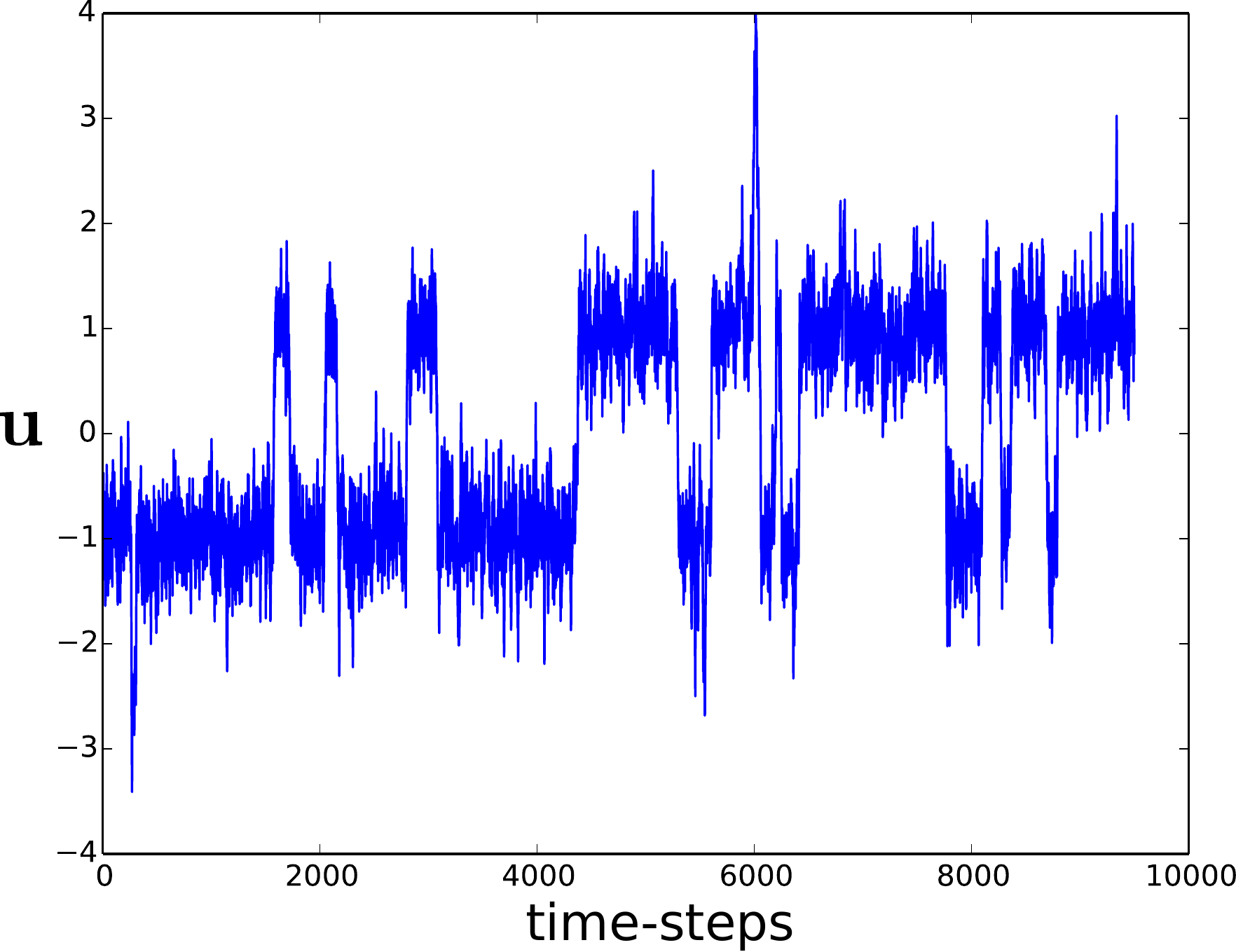}\label{fig: double well traj}}
        \caption{Illustrations of the double well process \eqref{eq: double well process}. (left) Energy landscape. (right) Trajectory of the process. The simulation was done using Euler-Maruyam algorithm with $\sigma = 0.7$, $dt = 0.1$ during $10000$ time steps. These parameters are kept constant for the next simulations.}
        \label{fig: double well}
\end{figure}

We consider a synthetic example where the data are generated as the solution of the stochastic differential equation corresponding to a particle in an energy landscape made of two different wells, as shown in Fig.\ref{fig: double well landscape}. More precisely, the target is a one-dimensional process described by
\begin{equation}
 d\u = -\nabla_\u E dt+  \sigma dB_t
 \label{eq: double well process}
\end{equation}
where $\nabla_\u E = \left\{
    \begin{array}{ll}
        1 & \mbox{if } -1<\u<0 \mbox{ or } \u>1\\
        -1 & \mbox{else}
    \end{array}
\right.$ is the gradient of the function described in Fig.\ref{fig: double well landscape} at point $\u$. The typical behavior of such a system is illustrated in Fig.\ref{fig: double well traj}. Roughly speaking, the particle jumps from one well to the other after random durations. Informally, each well can be said to be an attractor.

\begin{figure}[htbp]
        \centering
                \subfigure[Relative entropy vs number of neurons.]{\includegraphics[width=0.3\textwidth]{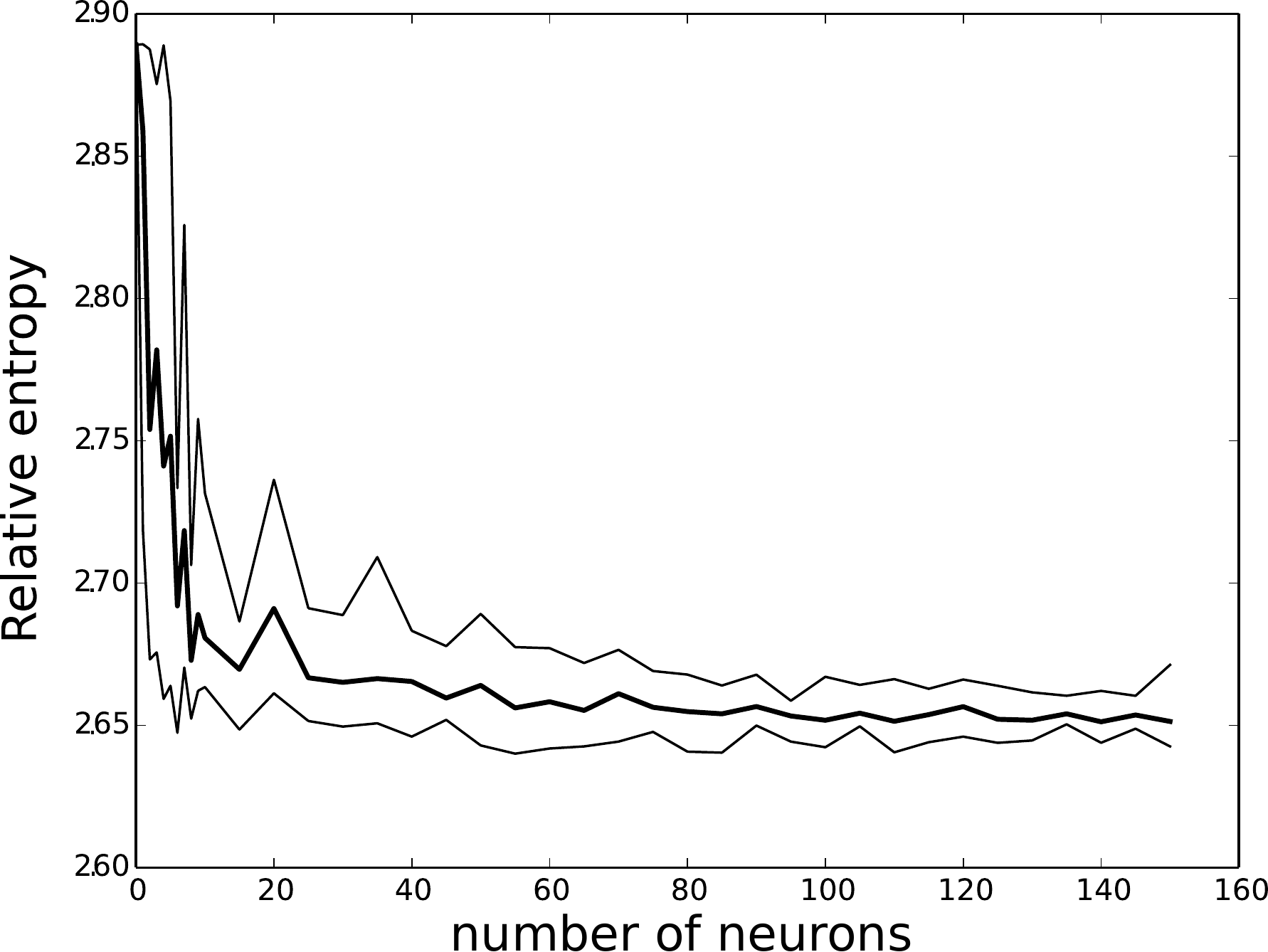}\label{fig: double well entrop vs nb neur}}\quad
                \subfigure[LIM trajectory.]{\includegraphics[width=0.3\textwidth]{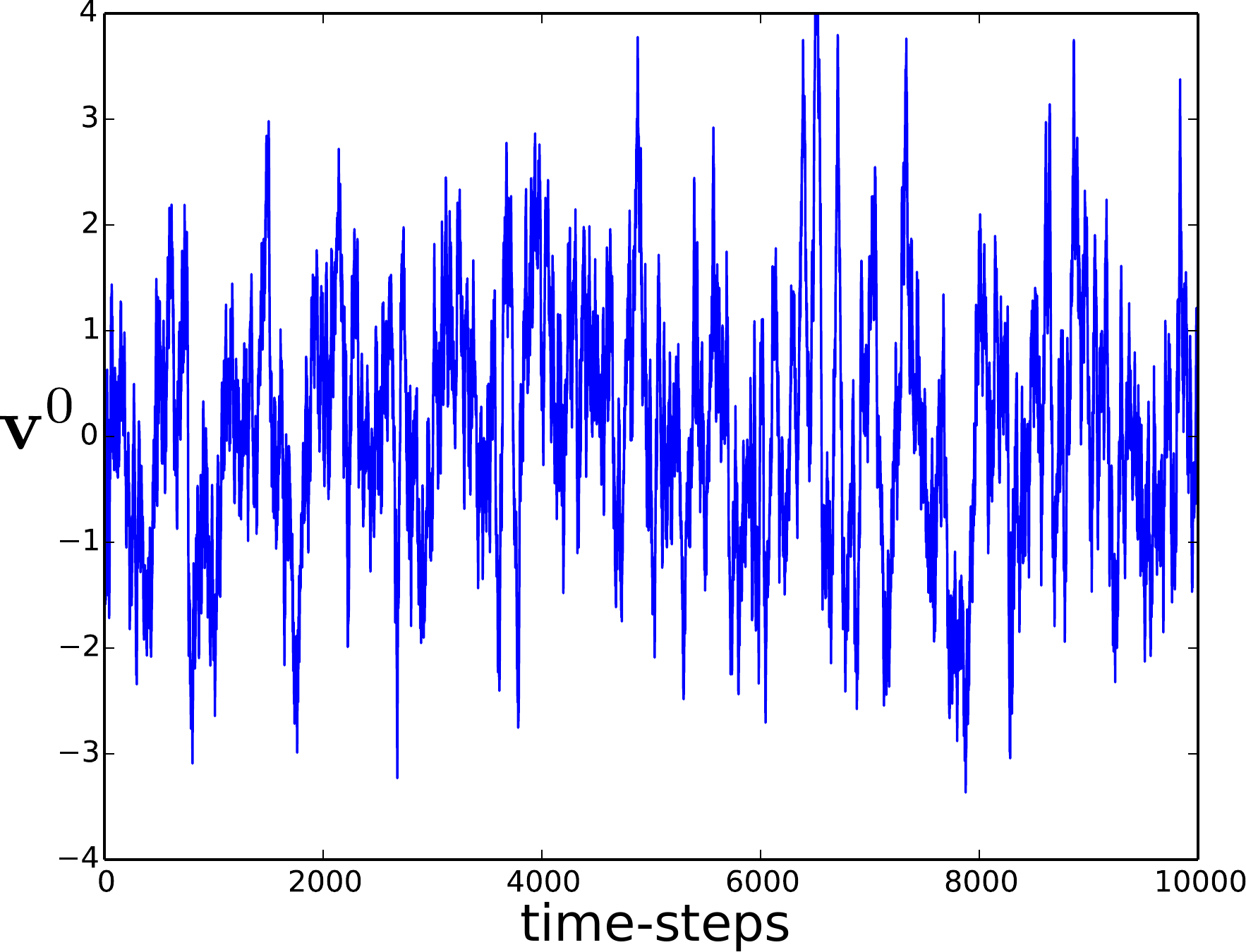}\label{fig: double well LIM traj}}\quad
                \subfigure[ESNsto trajectory.]{\includegraphics[width=0.3\textwidth]{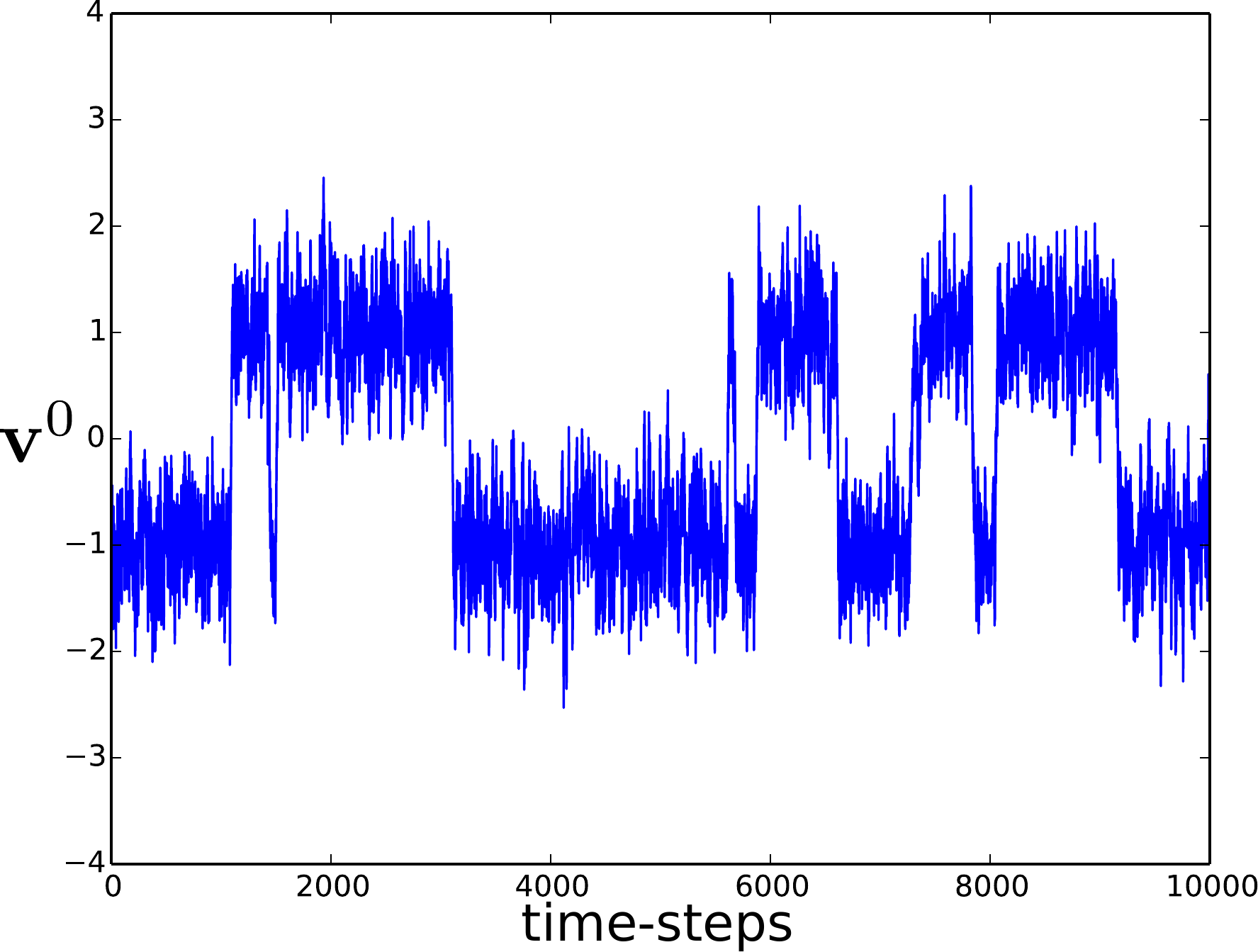}\label{fig: double well ESNsto traj}}                
        \caption{(a) Relative entropy as a function of the number
                of neurons in the reservoir. For any number of
                neurons, we launched $10$ simulations, to take into
                account the variability of reservoir connections, and
                measured the relative entropy according to equation
                \eqref{eq: relative entropy}. The top (resp. bottom)
                curve is the maximum (resp. minimum) value of the
                entropy among the different trials. The middle curve
                is the mean. The relative entropy roughly decreases
                from $2.9$ to $2.65$ which is enough to qualitatively
                change the behavior of the neural networks as shown in
                the left and middle picture. (b) Result of the
                LIM simulation (i.e. $m=0$) learned from the data in
                Fig.\ref{fig: double well traj}. (c) Result of the
                ESNsto simulation with $m=100$ neurons. Parameters
                used: $\eps = 1$, $l = 1$, $s(x) = \tanh(2x)$, $\Wcc$
                is drawn according to a normal law and rescaled so
                that $\|\Wcc\| = 1$ (operator norm), $\Wcr$ is drawn
                uniformly in $[-1.1]$, $\alpha = 1$. The network
                simulations and  
          learning were done during $T = 10^5$ time steps.
          Note that there is a washout time interval of $500$ at the
          beginning of the learning (and reconstruction) in order to
          remove transient effects.}
        \label{fig: double well reconstruction}
\end{figure}

We now compare the approximations of this time series based on LIM and
the ESNsto. We see in Fig.\ref{fig: double well entrop vs nb neur}
that increasing the number of neurons in the reservoir improves the
relative entropy defined by \eqref{eq: relative entropy}. With no
neurons in the reservoir, which corresponds to LIM, the relative
entropy is approximately $2.9$. After a sharp decrease for the first
dozens of additional neurons, the relative entropy slowly decreases
when the number of neurons increases to finally reach a value close to
$2.65$ for $m=150$ neurons. Note that the simple numerical
differentiation method that we have used imposes a lower bound on the
relative entropy shown here. Indeed, it is easy to observe that the
relative entropy between system \eqref{eq: double well process} and
itself using definition \eqref{eq: relative entropy} is $2.45$. We
believe that the gap between the ESNsto and the optimal value will
decrease with additional neurons in the network, but may not vanish
due to the over-fitting issue mentioned later. This suggests that
appropriately choosing the regularization parameter is crucial for
optimal accuracy. We can also observe that the variance of the
relative entropy value, corresponding to different random realizations
for the connections in the reservoir $\Wcc$ and $\Wcr$, decreases with
the number of neurons: due to better averaging, large reservoirs are
less dependent on the realization defining their weights.

Running the LIM and ESNsto networks post-learning shows different qualitative
behaviors as displayed in Fig. \ref{fig: double well LIM traj} and
\ref{fig: double well ESNsto traj}. As opposed to LIM, the ESNsto
reproduces patterns of noise-induced jumps between two attractors.

\begin{figure}[htbp]
        \centering
                \subfigure[State distribution for the data.]{\includegraphics[width=0.3\textwidth]{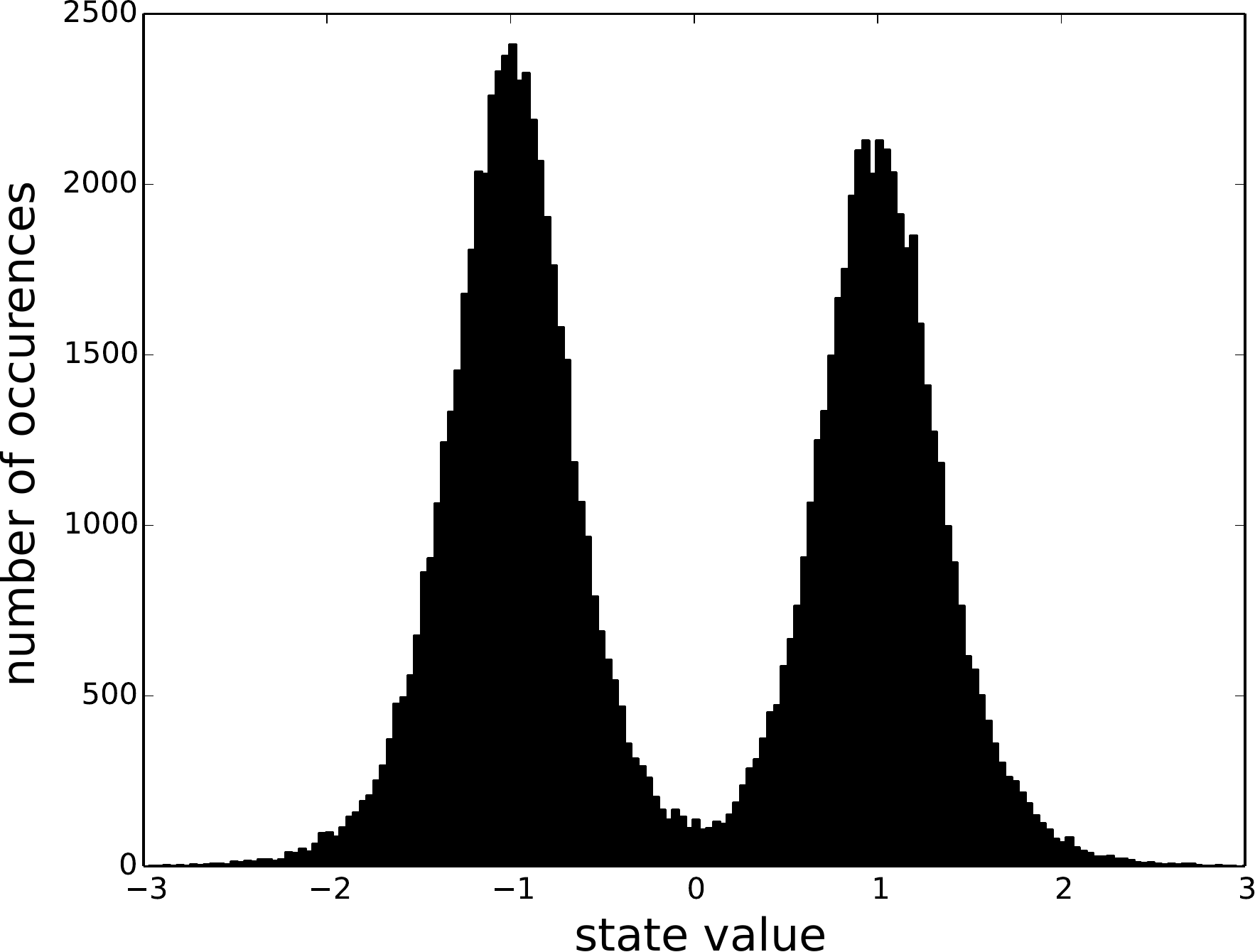}\label{fig: double well data distrib}}\quad
                \subfigure[State distribution for the LIM.]{\includegraphics[width=0.3\textwidth]{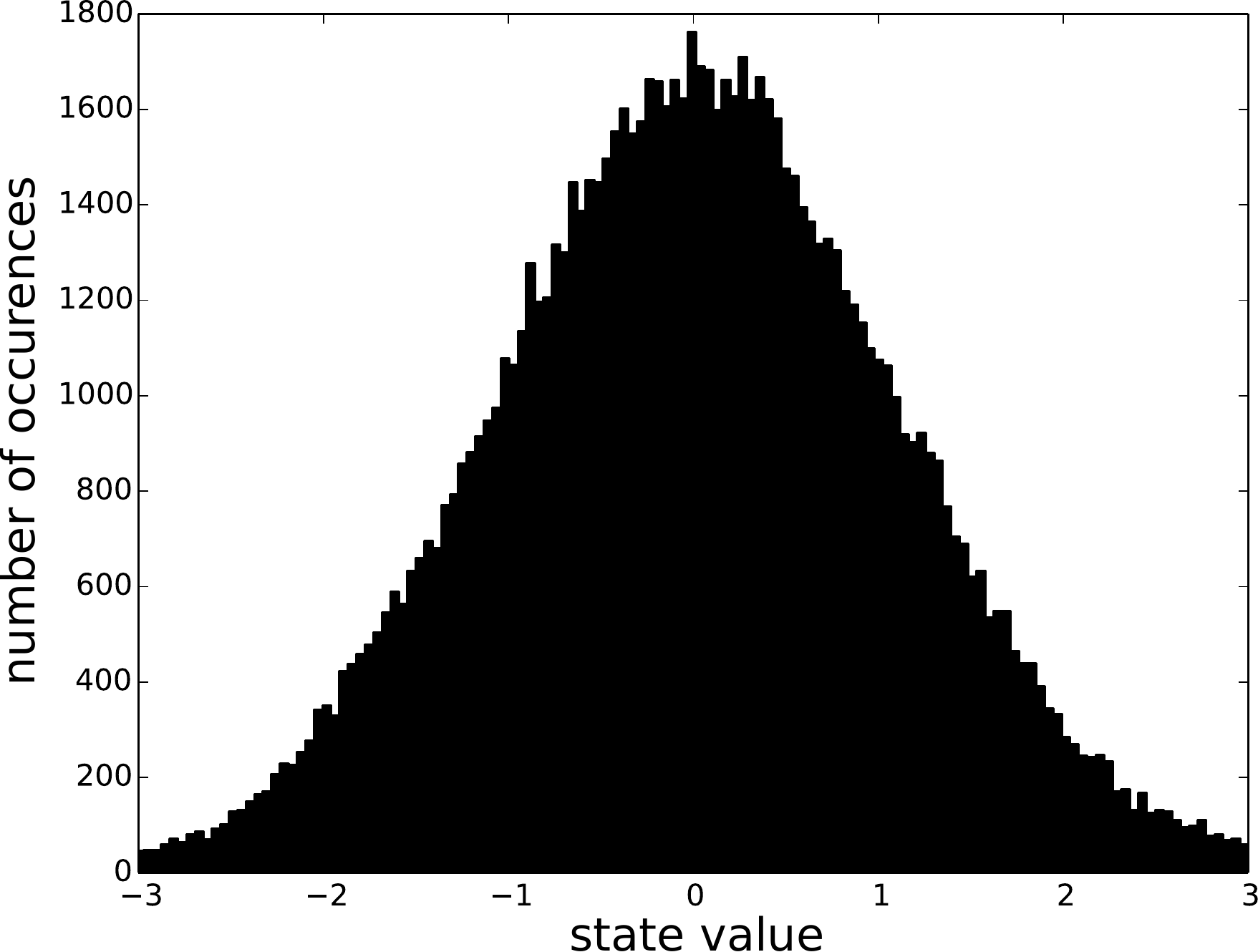}\label{fig: double well LIM distrib}}\quad
                \subfigure[State distribution for the ESNsto.]{\includegraphics[width=0.3\textwidth]{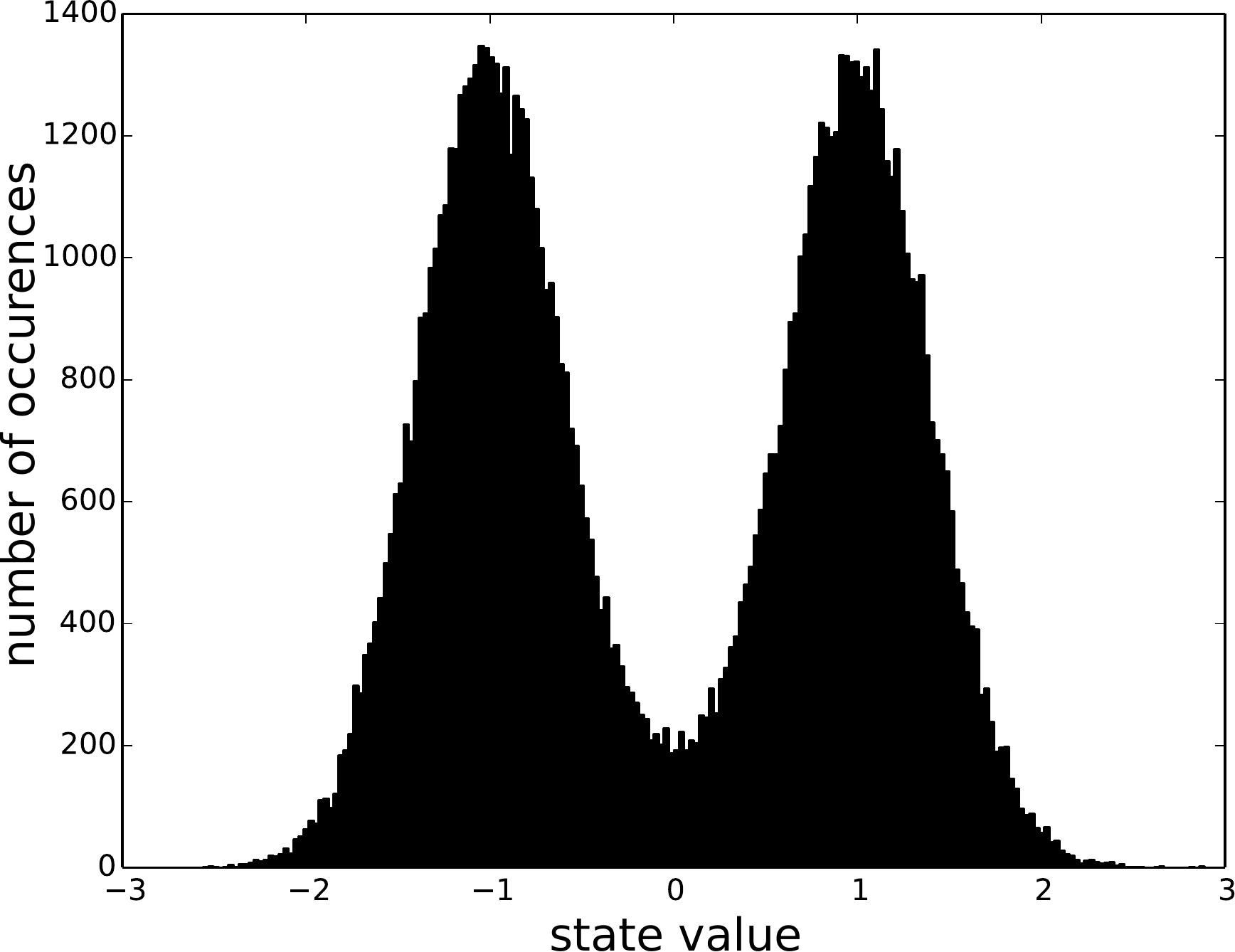}\label{fig: double well ESNsto distrib}}\\
                \subfigure[Distributions of the durations between jumps.]{\includegraphics[width=0.45\textwidth]{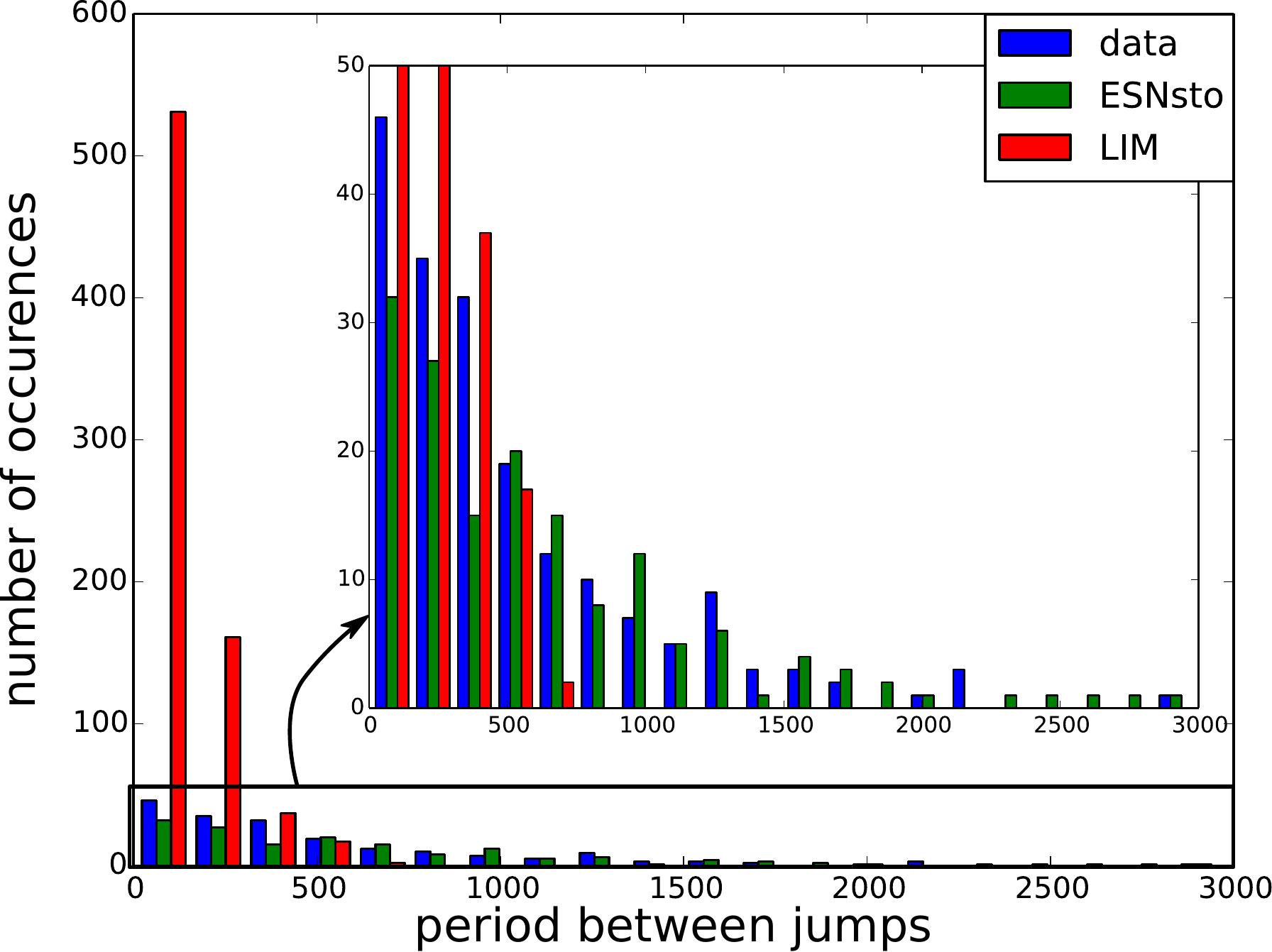}\label{fig: double well jump times}}\quad
        \caption{Distribution of the states of the data (a), LIM
                (b) and ESNsto (c). (d): Compared
                histograms of the duration spent in a well between two
                jumps. The parameters used are the same as for
                Fig.\ref{fig: double well reconstruction}.} 
        \label{fig: double well histograms}
\end{figure}

Significant qualitative improvements made possible by having neurons
in the reservoir can also be seen by empirically measuring some
statistical quantities of the target, LIM and ESNsto runs, as shown in
Fig.\ref{fig: double well histograms}. The first three figures
\ref{fig: double well data distrib}, \ref{fig: double well LIM
  distrib} and \ref{fig: double well ESNsto distrib}, show that LIM is
failing to reproduce the bimodal distribution of the target
corresponding to the two attractors: LIM has a
unimodal Gaussian-like distribution, whereas ESNsto is able to
reproduce the bimodality thanks to the non-linearities in the
reservoir.

In Fig. \ref{fig: double well jump times}, it is shown that the
distribution of the times spent in each attractors between two jumps
is irrelevant for LIM whereas it is similar between target and ESNsto.
This is also reflected in the transition rates which is approximately
$0.0019$ for the data, $0.0016$ for the ESNsto and $0.0075$ for the
LIM. However, it is to be noticed that an increase in the number of
neurons beyond 150 neurons leads to a decrease of the transition rate
(not shown). This underlines a drawback of the method for non-ergodic time series which exhibit significant noisy flucutuations: the network
tries to put in the connectivity as much variability as possible; the noise term is simply taking care of the left-overs. Therefore, the noise induced transitions in the target are
not only modeled by the diffusion term in the neural network but also
by the drift. This effect will vanish if the learning time series has enough ergodicity so that the noisy behavior is averaged out when computing the drift. When there is a limited amount of time steps available, a better numerical differentiation scheme may improve the approximation accuracy since it would filter out noise before asking the drift to approximate it.

\subsection{El Ni\~{n}o phenomenon}
\begin{figure}[ht]
 \centering
 \includegraphics[width=0.8\textwidth]{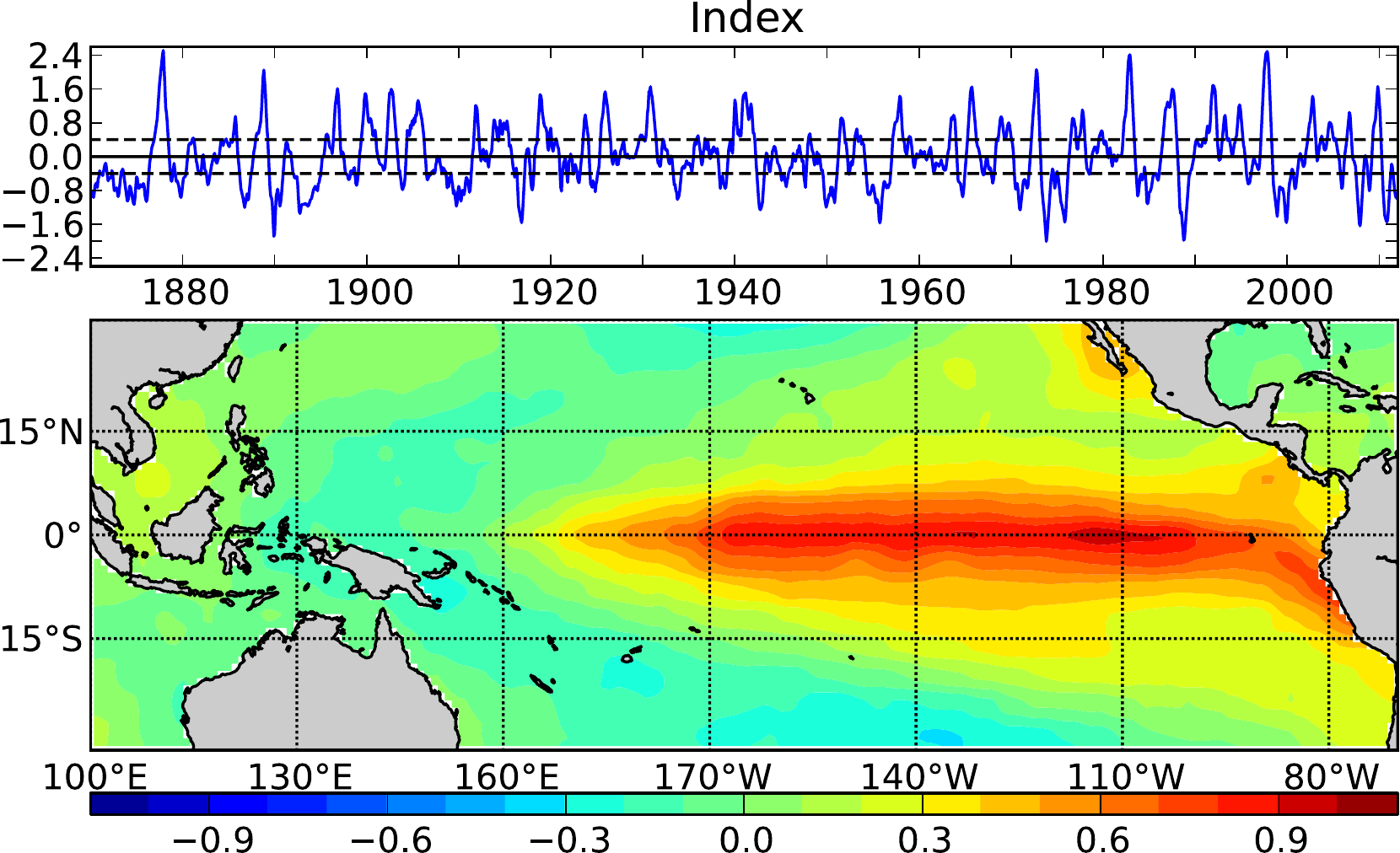}
 \caption{Ni\~{n}o 3.4 index, smoothed with a 3-month running mean (top) and associated SST anomalies (bottom).}
 \label{fig:N34index}
\end{figure}

In this section, we focus on approximating the geophysical process El
Ni\~{n}o. It corresponds to a large warming of Sea Surface Temperature
(SST) in the eastern equatorial Pacific, occurring irregularly every 2
to 7 years, and having a broad impact of the global climate
\cite{trenberth1997definition,deser2010sea}.  As many other
geophysical processes, its dynamics evolves on different interacting
timescales.  It is in particular strongly linked to atmospheric
processes evolving at shorter time scales with a nonlinear behaviour.
Penland showed that the evolution of 3-month running mean SST
anomalies in the tropical Indo-Pacific, and thus of El Ni\~{n}o, are
well approximated by a LIM, where the rapidly varying nonlinear
processes are parameterized as a stochastic forcing of the slower
system \cite{penland1996stochastic}. One commonly used index of the El
Ni\~{n}o phenomenon is the Ni\~{n}o 3.4 index (N34 index), defined as
the averaged of SST anomalies between 5\textdegree S- 5\textdegree N
and 170\textdegree W-120\textdegree W. Using the definition of
\cite{trenberth1997definition}, an El Ni\~{n}o event is said to occur
when the N34 index, smoothed with a 5-month running mean, exceed
0.4\textdegree C for 6 months or more. Fig. \ref{fig:N34index} shows
the N34 index (top) and the regression of SST anomalies onto this
index (bottom), indicating the warming in the eastern equatorial
Pacific associated with a positive N34 index.

The target time series considered here are the N34 index, smoothed
with a 3-month running mean, and the 10 first principal components
(PCs) of the empirical orthogonal function of 3-month running mean SST
anomalies in the IndoPacific region (30\textdegree S-30\textdegree N,
40\textdegree E-70\textdegree W). These 10 PCs represent 80\% of the
total variance of monthly IndoPacific SST anomalies, and they are used
instead of considering directly the SST anomalies at each grid point
in the IndoPacific to reduce the dimensionality of the system.  The
data come from the HadISST1 SST dataset \cite{rayner2003global},
constructed from in situ SST observations and satellite derived
estimates and available from 1870. We used data from 1870 to 2011. Our
target times series contain thus 1704 time steps (corresponding to
1704 months or 142 years) and a washout period of 240 time steps is
used at the beginning of the learning to remove transient effects.

\begin{figure}[htbp]
        \centering
                \subfigure[Ridge regularization parameter ranging from 0 to 5000]{\includegraphics[width=0.4\textwidth]{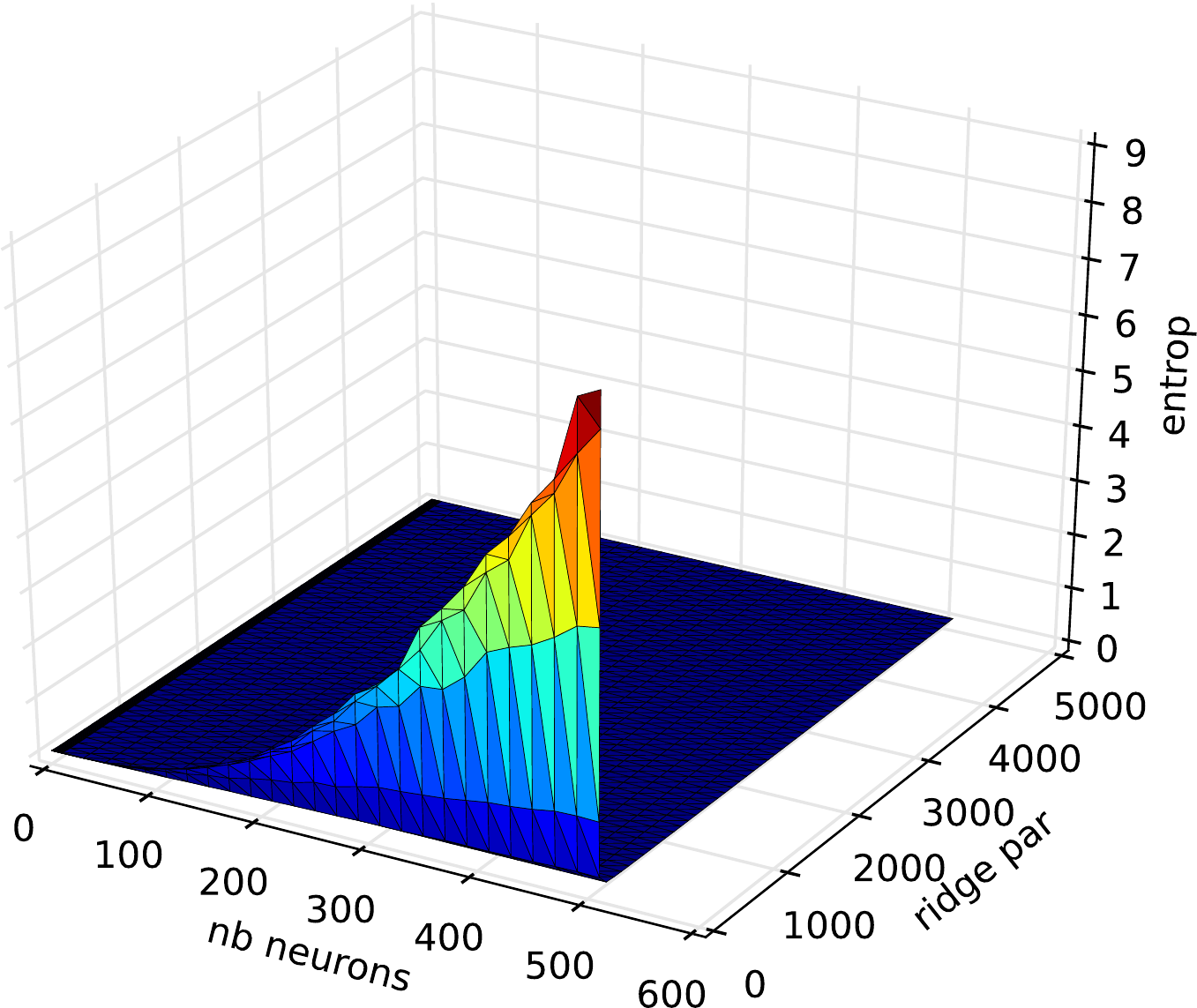}\label{fig:surf_entrop_N34_large}}\quad
                \subfigure[Zoom for ridge regularization parameter ranging from 500 to 5000]{\includegraphics[width=0.4\textwidth]{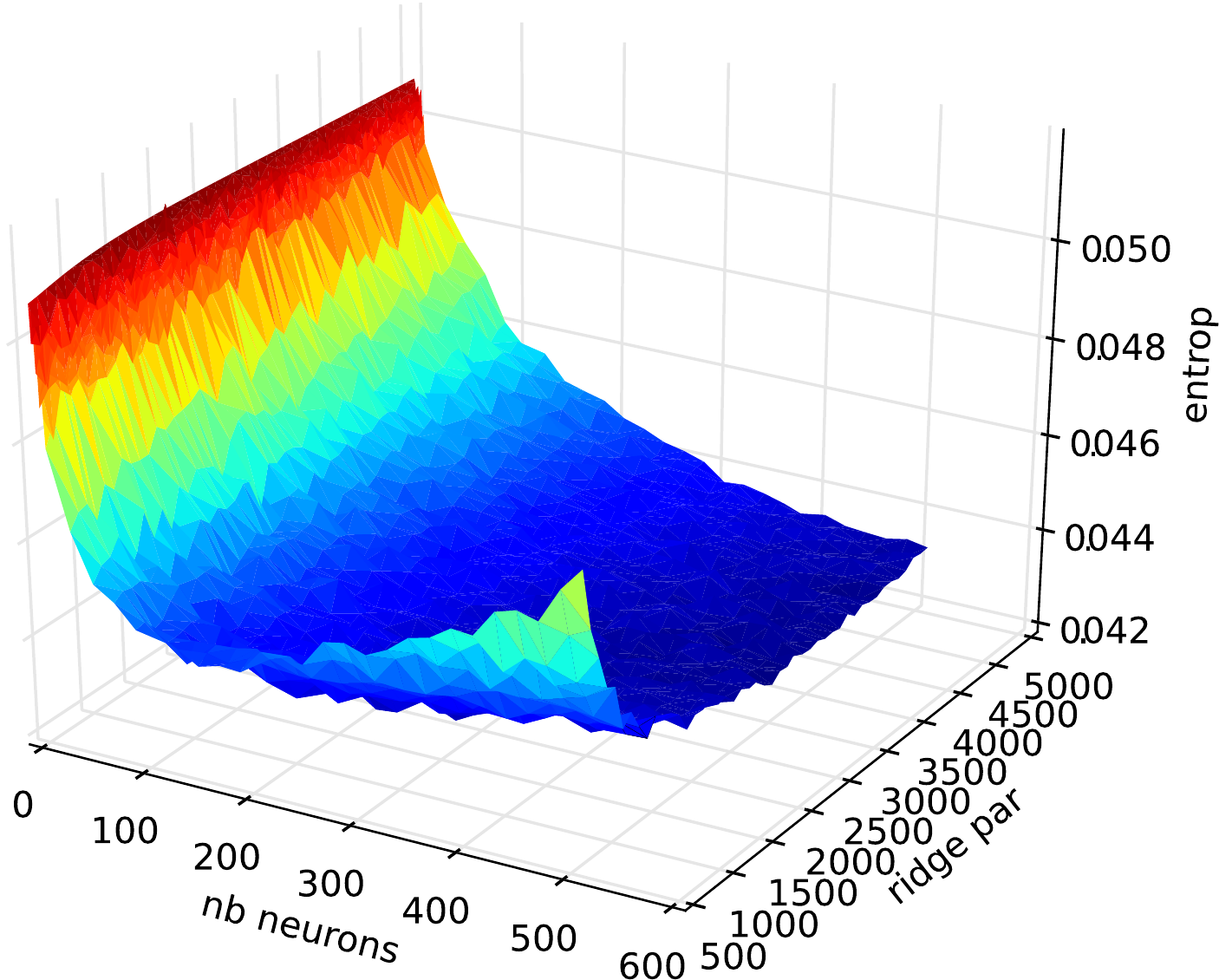}\label{fig:surf_entrop_N34_zoom}}
        \caption{Relative Entropy as a function of the number of neurons $m$ and the ridge regularisation parameter $\alpha$. Parameters used: $\eps = 0.1$, $l = 1/2$, $s(x) = \tanh(2x)$, $\Wcc$ is drawn according to a normal law and rescaled so that $\|\Wcc\| = 1$ (operator norm), $\Wcr$ is drawn uniformly in $[-1.1]$, $\alpha = 1$. Washout $=240$. The cross-validation was done with $k=10$ blocks. Each value of the relative entropy corresponds to the average over 10 realizations of the weight matrices.}
        \label{fig:surf_entrop_N34}
\end{figure}

Here, we compare the approximations of the N34 index based on LIM and
ESNsto. A crucial parameter for the success of ESNsto was the choice
of $\eps=0.1$ in equation \eqref{eq: activities}. It is known that
this parameter controls the speed of the reservoir
\cite{jaeger2007optimization}. In our case, the reservoir needed to
evolve, not at the scale of months (which would have corresponded to
$\eps=1$), but rather at longer time scales to be helpful in
reconstructing the dynamics. Fig. \ref{fig:surf_entrop_N34} shows the
relative entropy of the system as a function of the number of neurons
and the ridge regularization parameter $\alpha$. Without any
regularization, the relative entropy increases with the number of
neurons. However, for $\alpha>10$, the relative entropy decreases with
the number of neurons and with $\alpha$, suggesting that ESNsto leads
to a better approximation than LIM when using strong regularization.
This can be interpreted as overfitting in the case of
weak regularization. Adding regularization penalizes the
accuracy on the training dataset (not shown) but significantly improve the
generalization on the test dataset, as observed in
Fig. \ref{fig:surf_entrop_N34_zoom}.

We now compare the simulations based on LIM and ESNsto with $m=500$
and $\alpha=3000$. This choice of $m$ and $\alpha$ is motivated by
Fig. \ref{fig:surf_entrop_N34}. The system is simulated forward for
$10^{5}$ time steps, corresponding to more than 8000 years. 

\begin{figure}[htbp]
        \centering
                \subfigure[Spectrum of the N34 index. Frequency in cycle per month.]{\includegraphics[width=0.3\textwidth,height=0.25\textwidth]{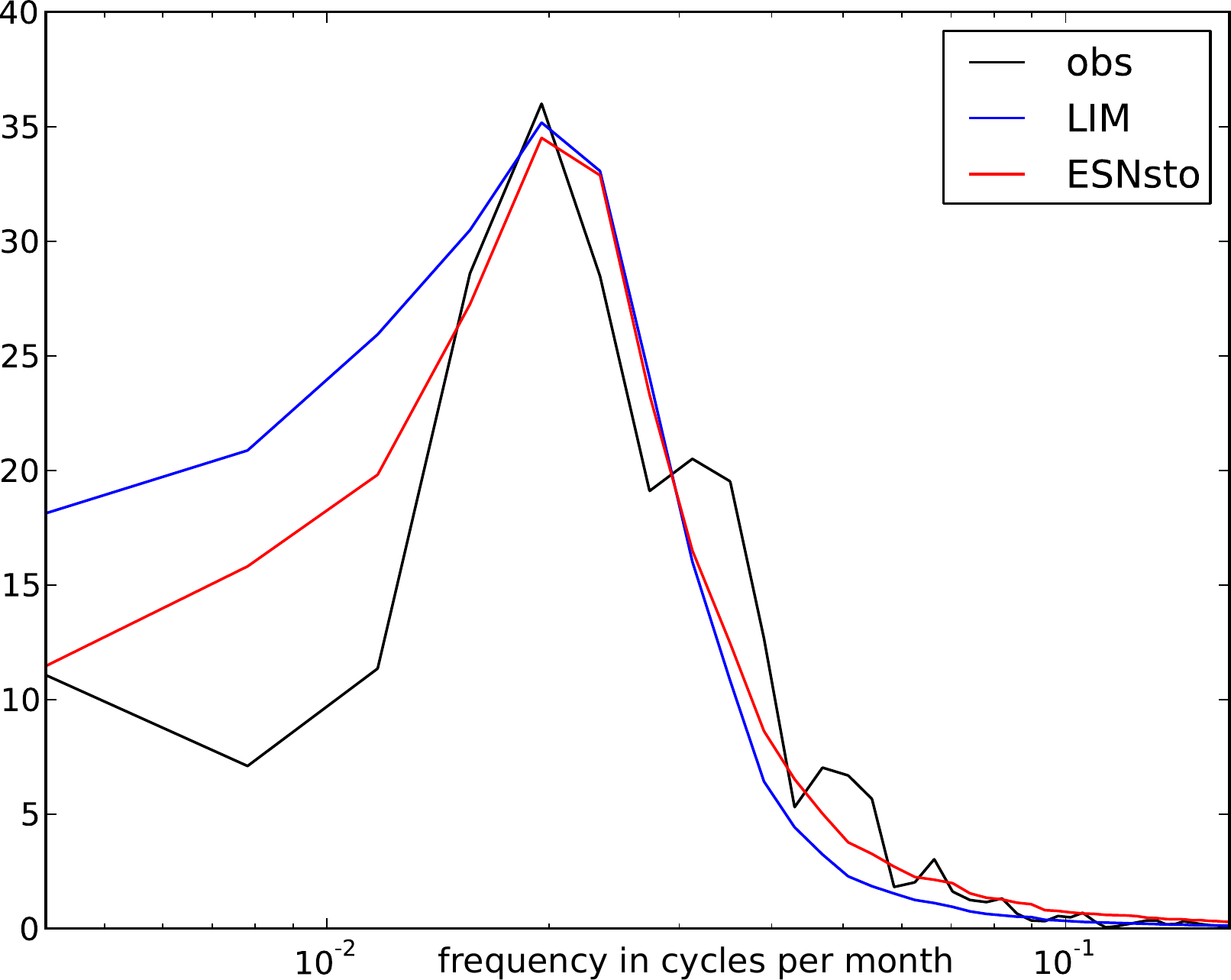}\label{fig:N34spectrum}}\quad
                \subfigure[Distribution of the N34 index]{\includegraphics[width=0.3\textwidth,height=0.25\textwidth]{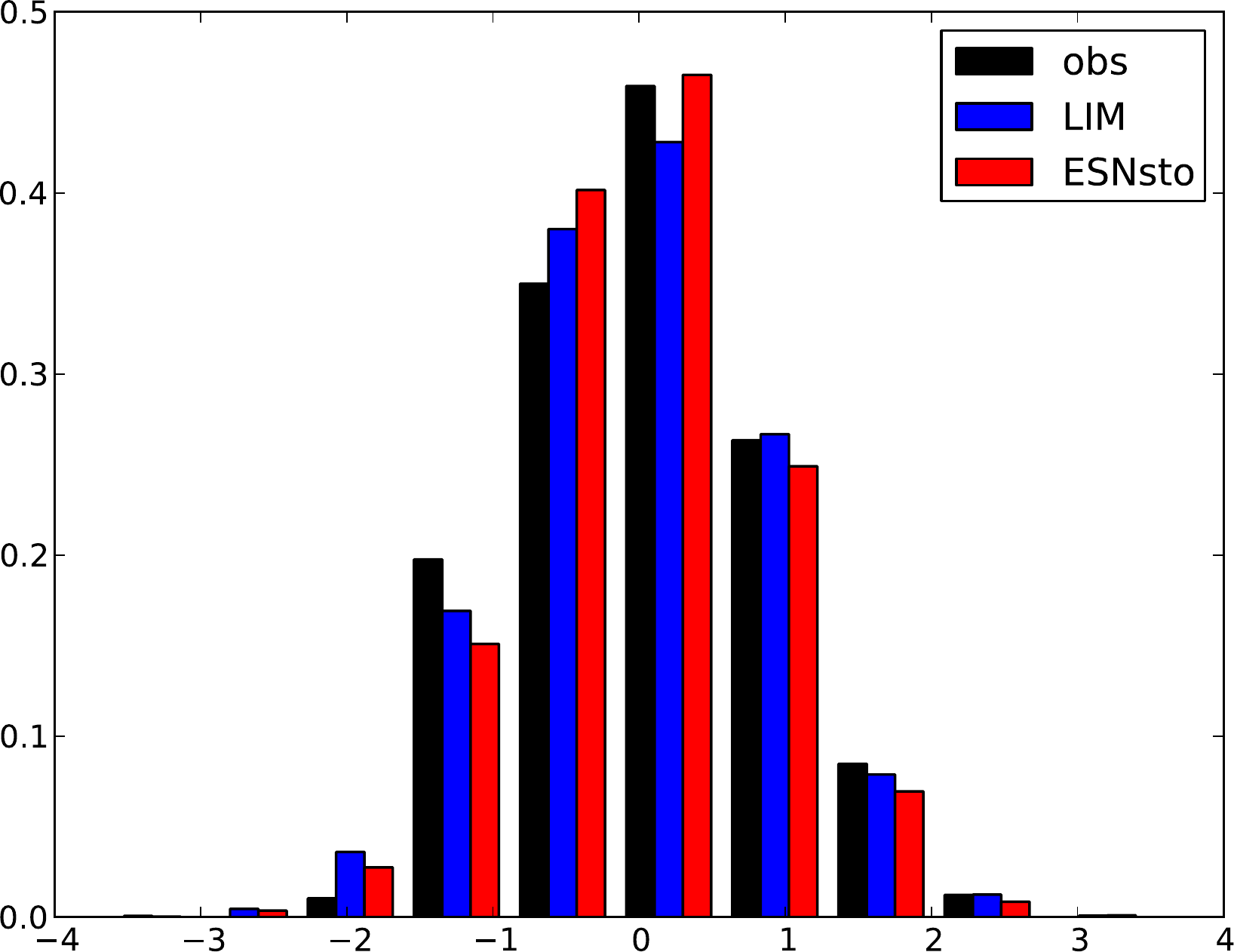}\label{fig:N34distrib}}\quad
                \subfigure[Distribution of the time interval between two El Ni\~{n}o events]{\includegraphics[width=0.3\textwidth,height=0.25\textwidth]{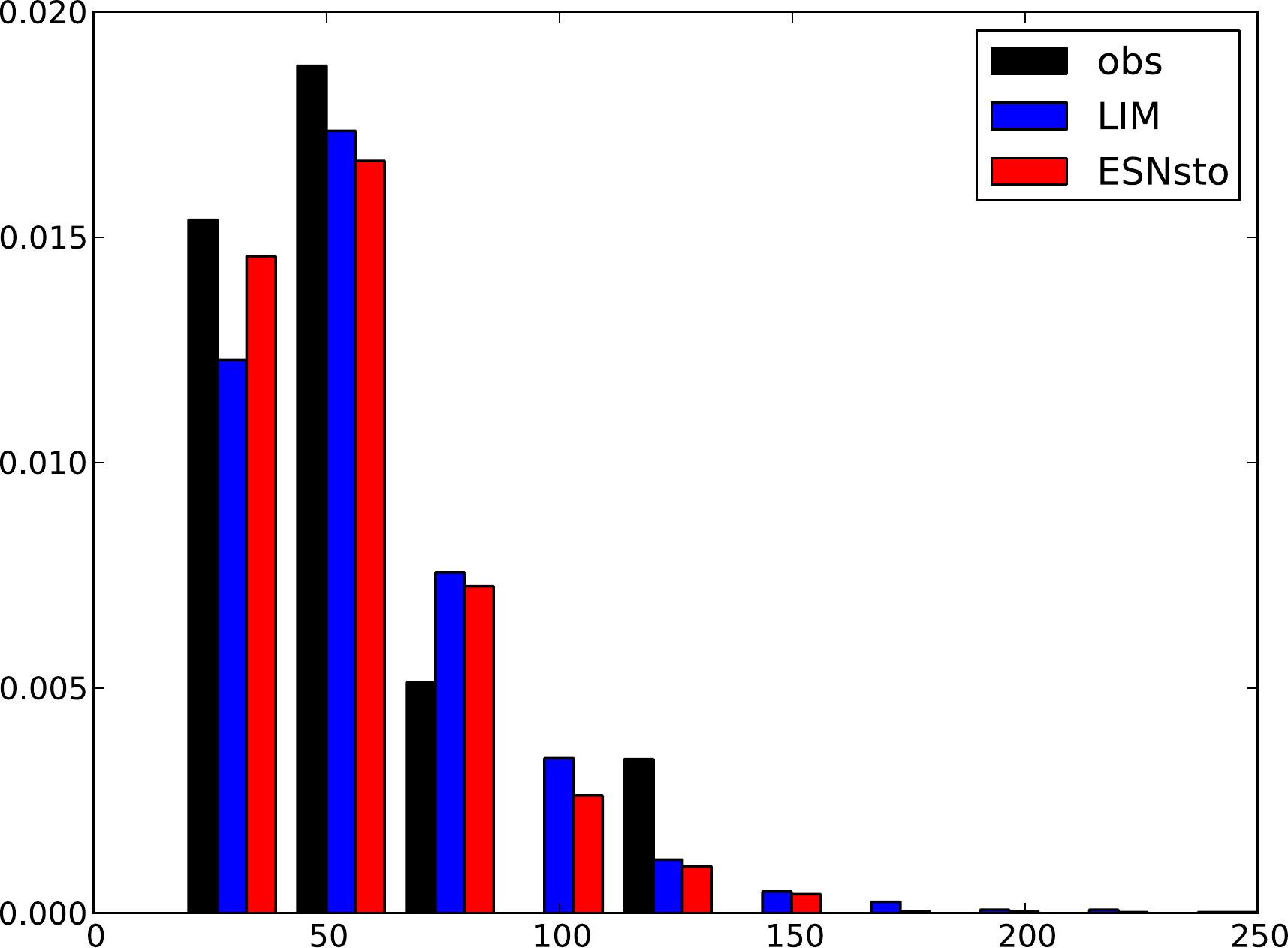}\label{fig:evN34distrib}}
        \caption{Spectrum (left) and distribution (middle) of the N34 index, smoothed with a 3-month running mean. Distribution of the time interval between two El Ni\~{n}o events (right). Black corresponds to the targeted time series, blue to LIM, and red to ESNsto. Parameters are identical to those of Fig. \eqref{fig:surf_entrop_N34}.}
        \label{fig:N34characteristics}
\end{figure}

As shown in Fig. \ref{fig:N34spectrum}, the spectrum of the N34 index
is closer to the target when approximated by ESNsto than by LIM.
At time scales longer than 4-5 years, the latter shows too much
variability. The spectrum obtained with ESNsto also shows a too high
variability, but less than the LIM and only at time scales longer than
7 years. 

The distributions of the N34 index based on the LIM and ESNsto
simulations are compared with the targeted distribution in Fig.
\ref{fig:N34distrib}. Due to the low number of target samples, it is
hard to determine from the figure which distribution is closer to the
target. A Kolmogorov-Smirnov test is used to determine whether the
simulated distributions differ from the targeted one. The p-values in
the case of LIM and ESNsto are respectively $0.45$ and $0.62$, meaning
that, in both cases, we cannot reject the null hypothesis that the
simulated and targetted N34 indices are drawn from the same
distribution. The larger p-value in the case of ESNsto indicates a
stronger evidence against the null hypothesis. 
 
Fig.\ref{fig:evN34distrib} shows the distributions of the time
interval between two El Ni\~{n}o events. Again, a Kolmogorov-Smirnov
test is used to estimate if the simulated distributions differ from
the targeted one, and gives a p-value of $0.76$ for LIM and $0.98$ for
ESNsto. The p-value almost equals 1 in the case of ESNsto, suggesting
a very good accuracy of our model to reproduce the some aspects of the
dynamics governing El Ni\~no events.

\section{Discussion}
We have shown how to design a recurrent neural network to reproduce a
target stochastic process. By doing so, we have introduced a rigorous
mathematical derivation which unifies ESNs and LIM under the general
principle of relative entropy minimization. Finally, we have shown how
the proposed system outperforms LIM, even when the parameters are
only coarsely tuned, on a simple synthetic task and on a climate
example of well-known importance.

We have observed that the system is prone to over-fitting, which
forced us to use large regularization parameters. Indeed, the
sequential computation of connectivity matrix followed by noise
matrix, implies that noise only takes care of left-overs. The
connectivity matrix will try to encode as much of the signal as
possible, even some part of the inherent noise. This is problematic in
applications where the target process is significantly noise driven.
However, when the number of time steps of the target time series is
large enough to have a good ergodic approximation or if we improve the
numerical differentiation scheme (e.g. using the Savitzky-Golay
algorithm), we believe this drawback will vanish.

Possible extensions of this theory could include a proper treatment of the case of badly sampled data as well as the generalization of the method to space dependent diffusion coefficients. Finally, an important step in increasing accuracy of these networks would be to identify an appropriate automatic tuning of the hyper parameters of the network (e.g. the spectral radius of the reservoir connections).

\section{Acknowledgment}
The authors would like to thank Mantas Lukosevicius for helpful discussions. MNG was funded by the Amarsi European Project.

\bibliographystyle{apalike}
   \bibliography{./info_reservoir_bib}
\end{document}